\documentstyle[aps,prd,twocolumn,amsfonts,epsfig,floats]{revtex}

\newcommand{\mb}[1]{\mbox{\boldmath $#1$}}

\begin{document} 
\draft \twocolumn[\hsize\textwidth\columnwidth\hsize\csname
@twocolumnfalse\endcsname

\title{Evolution of Cosmological Models in the Brane-world Scenario}

\author{Antonio Campos and Carlos F. Sopuerta}

\address{~}

\address{Relativity and Cosmology Group, School of Computer Science
and Mathematics, \\
Portsmouth University, Portsmouth~PO1~2EG, Britain}

\address{~}

\date{\today}

\maketitle


\begin{abstract}
In this work we consider Randall-Sundrum brane-world type scenarios, in 
which the spacetime is described by a five-dimensional manifold with matter 
fields confined in a domain wall or three-brane.  We present the results 
of a systematic analysis, using dynamical systems techniques, of the 
qualitative behaviour of the Friedmann-Lema\^{\i}tre-Robertson-Walker  
and the Bianchi I and V cosmological models in these scenarios.  We construct
the state spaces for these models and discuss how their structure changes
with respect to the general-relativistic case, in particular, what new
critical points appear and their nature, the occurrence of bifurcations and
the dynamics of anisotropy.
\end{abstract}
\vskip 1pc  ] 



\section{Introduction}\label{sec1}

String and membrane theories are promising candidates for
a unified theory of all forces and particles in Nature.
A consistent construction of a quantum string theory
is only possible in more than four spacetime dimensions.
Then, to make a direct connection of these theories with 
our familiar non-compact four-dimensional spacetime we are 
compelled to compactify the extra spatial dimensions to 
a finite size or, alternatively, find a mechanism to 
localize matter fields and gravity in a lower dimensional
submanifold.

Recently, Randall and Sundrum have shown that for
non-factorizable geometries in five dimensions there 
exists a single massless bound state confined in a
domain wall or three-brane \cite{RanSun:1999b}. 
This bound state is the zero mode of the Kaluza-Klein 
dimensional reduction and corresponds to the 
four-dimensional graviton.
The picture of this scenario is a five-dimensional
Anti-de Sitter space ({\it bulk}) with an embedded three-brane 
where matter fields are confined and Newtonian gravity is effectively 
reproduced at large-scale distances.
Earlier work on Kaluza-Klein dimensional reduction and matter
localization in a four-dimensional manifold of a 
higher-dimensional non-compact spacetime can be found in~\cite{LOWD}.

The Randall-Sundrum model was inspired by string theory.
In the context of dimensional reduction of eleven-dimensional 
supergravity, Ho$\check{\mbox{r}}$ava and Witten showed 
that the ten-dimensional $E_8\times E_8$ heterotic string 
is connected with an eleven-dimensional theory compactified 
on the orbifold ${\mathbb{R}}^{10}\times S^1/{\mathbb{Z}}_2$
\cite{HorWit:1996a}.
Moreover, they concluded that the coupling constants 
of gauge fields in the ten-dimensional boundary are 
related with the eleven-dimensional gravitational constant
\cite{HorWit:1996b}.    
The picture coming out of this model is that of
two separated ten-dimensional manifolds. 
Gauge fields are confined in these boundary manifolds
whereas gravity can propagate in the higher dimensional 
spacetime. As a consequence, these two separated worlds
can only communicate through gravitational interactions.
The cosmological implications of the 
Ho$\check{\mbox{r}}$ava-Witten theory have already
been extensively analyzed~\cite{HOWI}.

The original motivation for the Randall-Sundrum model 
was the solution of the hierarchy problem in a slightly 
different set up \cite{RanSun:1999a}. 
In this case one has two parallel branes with opposite 
tensions embedded in a five-dimensional spacetime with
negative cosmological constant. 
Actually, the fifth dimension is compactified in the 
orbifold $S^1/{\mathbb{Z}}_2$ and the two branes are 
located at the singular boundary points. 
Due to an exponential factor in the metric tensor, the
particles living in the negative tension brane acquire 
effectively a huge physical mass parameter compared to 
the fundamental scale with a moderate fine tuning of the
size of the extra dimension.  
Unfortunately, as pointed out in~\cite{UNFO},
the cosmology in this brane is rather unsatisfactory
because the energy density of matter present in the brane
must be negative, which violates the weak energy condition.
The argument is based on the observation made by Bin{\'e}truy 
{\it et al.} \cite{BinDefLan:2000,BinDefEllLan:2000}
(see also \cite{FlaTyeWas:2000b})
that the effective Friedmann equation for the Hubble parameter
for a five-dimensional spacetime with energy density localized
in a infinitely thin domain wall is modified with respect to 
the general relativistic case.
Other attempts to solve the hierarchy problem in the context
of extra dimension have been examined in~\cite{EXDI}. 

Nevertheless, the model with a non-compact fifth dimension and 
only one brane is consistent with present gravity experiments. 
In general, scenarios with extra dimensions predict corrections
to the Newtonian potential at short distances and important 
deviations from the standard evolution of the universe at early 
times. Then, current day cosmological observations, such as the
age of the universe or the abundances of light elements, 
cannot be used to constraint these models.
In contrast, the search for deviations of Newton's law is
their fundamental observational probe~\cite{ChuEveDav:2000}.
The fact that Newtonian gravity has been tested quite accurately up to
$1\, mm$ ($M_\ast\sim10^{-16}\, TeV$) limits the value of the 
fundamental scale associated with the five-dimensional
gravitational coupling constant:  
$M^{(5)}{}_P\, \gg \, M^{2/3}_PM^{1/3}_\ast \sim 10^5\, TeV$ 
\cite{RanSun:1999b,MaaWanBasHea:2000} ($M_P$ is the 
Planck mass). Future experiments will further constraint this
naive estimate \cite{LonChaPri:1999,HoyEotWas:2000}

The purpose of the present work is to study the cosmological 
evolution of these brane-world scenarios.
We are going to follow the geometric formulation and generalization
of the Randall-Sundrum scenario introduced in 
\cite{ShiMaeSas:2000,SasShiMae:2000,Maa:2000}.
The Einstein equations in the bulk can be
written in the following form\footnote{Upper-case Latin letters denote
coordinate indices in the bulk spacetime ($A,B,\ldots = 0,\ldots,4$) 
whereas lower-case Latin letters denote coordinate indices in the 
four-dimensional spacetime where matter is confined 
($a,b,\ldots=0,\ldots,3$).  We will use physical units in which $c=1$.}
\begin{equation}
   G^{(5)}_{AB} 
        = -\Lambda_{(5)}g^{(5)}_{AB}
          +\kappa^2_{(5)}T^{(5)}_{AB},
   \label{eeob}
\end{equation}
with
\begin{equation}
   T^{(5)}_{AB} 
        = \delta(\chi)
          \left[ -\lambda\,g_{AB}
                  +T_{AB}
          \right].
   \label{eeob2}
\end{equation}
In these expressions $\kappa_{(5)}$ is the five-dimensional 
gravitational coupling constant; $g^{(5)}_{AB}$, $G^{(5)}_{AB}$ 
and $\Lambda_{(5)}$ are the metric, Einstein tensor and the negative
cosmological constant of the bulk spacetime, respectively;
$T_{AB}$ is the matter energy-momentum tensor; the spacelike 
hypersurface $x^4\equiv\chi=0$ gives the brane world and 
$g_{AB}$ is its induced metric; finally, $\lambda$ is the tension 
of the brane, which must be assumed to be positive in order to 
recover conventional gravity on the brane.
Using the Gauss-Codacci equations relating the four- and
five-dimensional spacetimes, equations~(\ref{eeob}-\ref{eeob2}) 
lead to the following modification of the Einstein's equations of 
General Relativity on the brane~\cite{ShiMaeSas:2000,SasShiMae:2000}
\begin{equation}
G_{ab}=-\Lambda g_{ab}+\kappa^2 T_{ab}+
\kappa^4_{(5)} S_{ab} - E^{(5)}_{ab} \,, \label{mefe}
\end{equation}
where $g_{ab}$ is the four-dimensional metric on the brane and 
$G_{ab}$ its Einstein tensor. The four-dimensional gravitational constant 
$\kappa$ and the cosmological constant $\Lambda$ are given in terms 
of the fundamental constants in the bulk by
\[ \kappa^2 = \frac{1}{6}\;\lambda\kappa^4_{(5)} \, ,\]
\[ \Lambda = \frac{|\Lambda_{(5)}|}{2}\;\left[\left(
\frac{\lambda}{\lambda_c}\right)^2-1\right] \, ,\]
respectively; where $\lambda_c$ is a critical brane tension\footnote{The 
particular Randall-Sundrum solution corresponds to the case when the 
tension of the brane $\lambda$ equals the critical brane tension 
$\lambda_c$~(\ref{cbt}) and $T_{ab}=E^{(5)}_{ab}=0$.}
given by
\begin{equation}
   \lambda^2_c
        = 6\; \frac{|\Lambda_{(5)}|}{\kappa^4_{(5)}}\,.
   \label{cbt}
\end{equation}
$S_{ab}$ are corrections quadratic in the matter variables (due to the
form of the Gauss-Codacci equations) and given by
\begin{equation}
S_{ab} = \textstyle{1\over12}T T_{ab}-\textstyle{1\over4}
T_a{}^c T_{bc}+\textstyle{1\over24}g_{ab}\left[3T^{cd}
T_{cd}-T^2\right] \,, \label{cobu}
\end{equation}
where $T\equiv T_a{}^a$.  And finally, $E^{(5)}_{ab}$ are corrections
coming from the extra dimension, more precisely, $E^{(5)}_{ab}$ are
the components of the {\em electric part} of the Weyl tensor of the bulk, 
$C^{(5)}_{ABCD}$, with respect to the normal, $n_A$ ($n^An_A=1$), to
the hypersurface $\chi=0$ where matter is confined, that is,
\[ E^{(5)}_{AB} = C^{(5)}_{ACBD}n^Cn^D \,. \]
Moreover, it is worth to note that the twice contracted second
Bianchi identities in the bulk, $\nabla_{(5)}^A\,G^{(5)}_{AB}=0$,
imply 
\begin{equation} 
\left. \nabla_a T^a{}_b \right|_{\chi=0} = 0 \,, \label{emtc}
\end{equation}
where we have taken $\{\chi,x^a\}$ to be Gaussian normal coordinates (see,
e.g.,~\cite{Wal:1984}) adapted to the hypersurface $\chi=0$.  Therefore,
we can say that the Einstein equations in the bulk~(\ref{eeob})
imply the conservation of the energy-momentum tensor in the brane world.

In this paper we will deal with generalized Randall-Sundrum 
scenarios in which the effects of the extra-dimension come
from the term quadratic in the energy-momentum tensor, i.e.
$S_{ab}$~(\ref{cobu}). Thus, we are assuming 
\begin{equation} 
E^{(5)}_{AB}|_{\chi=0} = 0~~\Longleftrightarrow~~E^{(5)}_{ab} = 0\,. 
\label{assu}
\end{equation} 
This includes conformally-flat
bulks ($C_{ABCD}=0$), and in particular, the five-dimensional 
Anti-de Sitter spacetime, the bulk considered in the original 
Randall-Sundrum scenario. The extension of this work to general 
bulks will be presented in a future paper~\cite{CamSop:2000}.

For the scenarios just outlined we have constructed and studied 
the state space of the Friedmann-Lema\^{\i}tre-Robertson-Walker 
(FLRW) and the Bianchi I and V cosmological models.  Then, we
have discussed systematically how the extra dimension changes the 
dynamics with respect to the general-relativistic case.  In particular, 
we find a new critical point representing the dynamics at very high 
energies, in the early universe (near the Big-Bang) and also 
near the Big-Crunch in the case of recollapsing models.  
We also find new bifurcations in the state space as the
equation of state of matter changes (we will assume a perfect-fluid
energy-momentum content), which are characterized by the 
occurrence of an infinite number of non-general-relativistic
critical points.  Finally, the Bianchi I and V models will 
provide information regarding the dynamics of anisotropy in
the brane-world scenario.

The paper is organized as follows. In section~\ref{sec2} we will
study the dynamics of the FLRW models in the brane-world scenarios,
introducing the notation and some tools used in the analysis of
dynamical systems~(see, e.g.,~\cite{DSCO,DSTH}).
In section~\ref{sec3}, we will study the dynamics in homogeneous but
anisotropic cosmological models.  In particular, the dynamics of the 
orthogonal Bianchi I and V cosmological models, which contain the 
flat and the negatively curved FLRW models, respectively.
We will finish with some concluding remarks in section~\ref{core}.



\section{Dynamics of the FLRW models in the brane-world scenario}
\label{sec2} 
In this section, we start by assuming that the brane-world is described
by a FLRW metric. The FLRW spacetimes are the standard cosmological
models. As is well-known~\cite{HawEll:1973,REGE}, 
they are motivated by the so-called {\em Cosmological Principle} in the 
sense that they are homogeneous and isotropic cosmological models (they
have a six-dimensional group of motions).  Then, the line element in
the brane-world ($\chi=0$) will be given by
\[ \mbox{ds}^2 = -dt^2 + a^2(t)\left[ dr^2+\Sigma^2_k(r)(
 d\theta^2+\sin^2\theta d\varphi^2)\right] \,, \] 
where
\[ \Sigma_k(r) = \left\{ \begin{array}{ll}
\sin r & \mbox{for $k=1$}\,, \\
r & \mbox{for $k=0$}\,, \\
\sinh r & \mbox{for $k=-1$}\,, \end{array} \right. \]
and $a(t)$ is the scale factor.  

Here, we will study the dynamics of the FLRW models considering a
bulk spacetime satisfying the condition~(\ref{assu}), 
which includes the five-dimensional Anti-de Sitter spacetime.  
On the other hand, we will assume that the matter content is equivalent
to that of a perfect fluid and therefore, the energy-momentum tensor will 
have the following form
\[ T_{ab} = (\rho+p)u_au_b + pg_{ab} \,, \]
where $\mb{u}$, $\rho$ and $p$ are the unit fluid velocity of matter 
($u^au_a=-1$), the energy density and the pressure of the matter fluid 
respectively.   We will also assume a linear barotropic
equation of state for the fluid, that is,
\begin{equation} 
p = (\gamma-1)\rho\,. \label{leoe}
\end{equation}
The weak energy condition (see, e.g.,~\cite{HawEll:1973}) imposes the
restriction $\rho\geq 0$, and from causality requirements, the speed of 
sound [$c_s\equiv (dp/d\rho)^{1/2}$] must be less than the speed of 
light, we have that $\gamma\in[0,2]$.
Then, taking into account the form of the equations~(\ref{mefe},\ref{cobu})
and~(\ref{assu}), it turns out that the fluid velocity $\mb{u}$  is aligned
with the velocity of the preferred observers in the FLRW spacetimes
(excepting in the case $G_{ab}\propto g_{ab}$, where there are no 
preferred observers), those that observe the matter distribution to 
be homogeneous and isotropic.  Then, we can write $\mb{u}$ as follows
\[ \vec{\mb{u}} = \frac{\mb{\partial}}{\mb{\partial t}} ~~
\Longrightarrow ~~ \mb{u} = - \mb{dt} \,. \]
Finally, taking into account recent observations~\cite{Rie:1998,Per:1999}, 
we will consider only the case of a positive cosmological constant, i.e. 
$\Lambda\geq 0$.   
Then, introducing the Hubble function $H(t)$
\[ H(t) \equiv \frac{1}{a}\frac{da}{dt}\equiv\frac{\dot{a}}{a}\,,\]
the dynamics of the FLRW models imposed by the modified Einstein field 
equations~(\ref{mefe}) and the energy-momentum conservation 
equation~(\ref{emtc}) is governed by the following set of ordinary 
differential equations
\begin{equation}
\dot{H}=-H^2-\frac{3\gamma-2}{6}\kappa^2\rho\left[1+\frac{3\gamma-1}
{3\gamma-2}\frac{\rho}{\lambda}\right]+\frac{1}{3}\Lambda\,,\label{raye}
\end{equation}
\begin{equation}
\dot{\rho}=-3\gamma H\rho\,, \label{emce}
\end{equation}
\begin{equation}
H^2=\frac{1}{3}\kappa^2\rho\left(1+\frac{\rho}{2\lambda}\right)
-\frac{1}{6}{}^3R+\frac{1}{3}\Lambda \,, \label{frie}
\end{equation}
where ${}^3R$ denotes the scalar curvature of the hypersurfaces
orthogonal to the fluid velocity, the $\{t=\mbox{constant}\}$ 
hypersurfaces, which is given by ${}^3R=6ka^{-2}(t)$.  
Equation~(\ref{raye}) is the
modified Raychaudhuri equation, equation~(\ref{emce}) comes from the
energy-momentum tensor conservation equation, and finally,
equation~(\ref{frie}) is the modified Friedmann equation. 
As is well-known, (\ref{raye}) is a consequence of (\ref{emce}) and 
(\ref{frie}), and the dynamics is completely described by the
functions $(H,\rho)$ and the parameters $k$, $\kappa$, 
$\gamma$, $\Lambda$ and $\lambda$.

In order to study the dynamics of these models we will closely follow 
the analysis carried out by Goliath \& Ellis~\cite{GolEll:1999}
for general relativistic FLRW models with a cosmological constant.
To that end, and in order to get compactified state spaces, it is 
convenient to consider two differentiated cases: 
(i) ${}^3 R \leq 0$ ($k=0$ or $k=-1$) and (ii) ${}^3 R > 0$ ($k=1$).

In the case (i), let us introduce the following set of dimensionless 
variables
\begin{eqnarray}
\Omega_\rho\equiv \frac{\kappa^2\rho}{3H^2}\,,  & ~~ &
\Omega_k\equiv -\frac{{}^3R}{6H^2}=-\frac{k}{\dot{a}^2}\,, \label{dlv1}
\\
\Omega_\Lambda\equiv \frac{\Lambda}{3H^2}\,, & ~~ &
\Omega_\lambda\equiv \frac{1}{6\lambda}\frac{\kappa^2\rho^2}{H^2} 
\,, \label{dlv2}
\end{eqnarray}
where $\Omega_\rho$ is the ordinary density parameter and 
$\Omega_k$, $\Omega_\Lambda$ and $\Omega_\lambda$ are the fractional
contributions of the curvature, cosmological constant and brane tension,
respectively, to the universe expansion~(\ref{frie}).  Therefore, all of 
them have a clear physical meaning.  As we can see, they are
non-negative and singular when $H=0$.  Furthermore, the Friedmann 
equation~(\ref{frie}), which now takes the following simple form
\begin{equation} 
\Omega_\rho+\Omega_k+\Omega_\Lambda+\Omega_\lambda=1\,.\label{fri2}
\end{equation}
implies that they must belong to the interval $[0,1]$ and hence, the 
state space with coordinates $\mb{\Omega}
=(\Omega_\rho,\Omega_k,\Omega_\Lambda,\Omega_\lambda)$ is compact.  

In order to find the dynamical equations for these variables we will
introduce the following dimensionless time derivative
\begin{equation}
' \equiv \frac{1}{|H|}\frac{d}{dt} \,, \label{tdkn}
\end{equation}
where $|H|$ is the absolute value of $H$.  Then, we have
\begin{equation}
H' = -\epsilon (1+q)H\,, \label{dece}
\end{equation}
where $\epsilon$ is the sign of $H$ [$\epsilon\equiv\mbox{sgn}(H)$].  
As is clear, for $\epsilon=1$ the model will be
in expansion, and for $\epsilon=-1$ it will be in contraction.
Moreover, $q$ is the deceleration parameter, which is defined by
\[ q\equiv -\frac{1}{H^2}\frac{\ddot{a}}{a} = \frac{3\gamma-2}{2}
\Omega_\rho-\Omega_\Lambda+(3\gamma-1)\Omega_\lambda \,. \]
Then, the dynamical system for our dimensionless 
variables~(\ref{dlv1},\ref{dlv2}) can be written in the following form
\begin{eqnarray}
\Omega'_\rho & = & \epsilon\left[2(1+q)-3\gamma\right]\Omega_\rho \,,
\label{orp1} \\
\Omega'_k & = & 2\epsilon q\Omega_k \,, \label{okp1} \\
\Omega'_\Lambda & = & 2\epsilon(1+q)\Omega_\Lambda \,, \label{olp1} \\
\Omega'_\lambda & = & 2\epsilon\left(1+q-3\gamma\right)\Omega_\lambda 
\,. \label{oap1}
\end{eqnarray}
It is important to note that equation~(\ref{dece}) is not coupled
to the system of equations~(\ref{orp1}-\ref{oap1}), and therefore we can
ignore it for the dynamical analysis.  To begin with, we have to find
the critical points of this dynamical system, which can be written in
vector form as follows
\[ \mb{\Omega}' = \mb{f}(\mb{\Omega}) \,, \]
where $\mb{f}$ can be extracted from~(\ref{orp1}-\ref{oap1}).  The critical 
points, $\mb{\Omega}^\ast$, which are the points at which the system 
will stay if initially it was there (see, e.g.~\cite{DSTH}), are given 
by the condition
\[ \mb{f}(\mb{\Omega}^\ast) = \mb{0} \,. \]
Their dynamical character is determined by the eigenvalues of the
matrix
\[ \left. \frac{\mb{\partial\mb{f}}}{\mb{\partial\Omega}}
\right|_{\mb{\Omega}=\mb{\Omega}^\ast} \,. \]
If the real part of the eigenvalues of a critical point is not zero, the 
point is said to be {\em hyperbolic}.  In this case, the dynamical 
character of the critical point is determined by the sign of the real
part of the eigenvalues:  If all of them are positive, the point is
said to be a {\em repeller}, because arbitrarily small deviations from this 
point will move the system away from this state.  If all of them are 
negative the point is called an {\em attractor} because if we move 
the system slightly from this point in an arbitrary way, it will return 
to it.  Otherwise, we say the critical point is a {\em saddle} point.
The dynamical system~(\ref{orp1}-\ref{oap1}) has four 
hyperbolic critical points corresponding to: 
the flat FLRW models ($\mbox{F}$), $k=\Lambda=\lambda^{-1}=0$
and $a(t)=t^{2/(3\gamma)}$;
the Milne universe ($\mbox{M}$), $\rho=\Lambda=0\,,$ $k=-1$ and $a(t)=t$; 
the de Sitter model ($\mbox{dS}$), $k=\rho=0$ and 
$a(t)=\exp(\sqrt{\Lambda/3}\,t)$; and a non-general-relativistic 
model ($\mbox{m}$) first discussed by Bin\'etruy, Deffayet and 
Langlois~\cite{BinDefLan:2000} in a brane-world scenario without brane
tension (see~\cite{BinDefEllLan:2000,FlaTyeWas:2000b} for more details).  
Their coordinates in the state space, i.e., 
$\mb{\Omega}=(\Omega_\rho,\Omega_k,\Omega_\Lambda,\Omega_\lambda)$, and 
their eigenvalues are given in the following table~\cite{Nota} 
\begin{quasitable}
\begin{tabular}{ccc}
Model & Coordinates   & Eigenvalues  \\  \tableline 
$\mbox{F}_\epsilon$ & $(1,0,0,0)$ & $\epsilon(3\gamma-2,3\gamma-2,
3\gamma,-3\gamma)$ \\
$\mbox{M}_\epsilon$ & $(0,1,0,0)$ & $\epsilon(-(3\gamma-2),0,2,
-2(3\gamma-1))$ \\
$\mbox{dS}_\epsilon$ & $(0,0,1,0)$ & $-\epsilon(3\gamma,2,2,6\gamma)$  \\
$\mbox{m}_\epsilon$ & $(0,0,0,1)$ & $\epsilon(3\gamma,2(3\gamma-1),6\gamma,
2(3\gamma-1))$
\end{tabular}
\end{quasitable}
The dynamical character of these points is given in a table below.

Now, let us consider the situation of the case (ii), in which ${}^3R$ is 
positive.  As we have already mentioned, in this case the state space defined 
by the variables $\mb{\Omega}
=(\Omega_\rho,\Omega_k,\Omega_\Lambda,\Omega_\lambda)$ is no longer compact
(because now $\Omega_k<0$).  However, we can introduce another set of 
variables, analogous to the ones introduced 
previously~(\ref{dlv1},\ref{dlv2}),
describing a compact state space.  Firstly, instead of using 
the Hubble function $H$ we will use the following quantity
\begin{equation}
D \equiv \sqrt{H^2+\textstyle{1\over6}{}^3R}\,, \label{vard}
\end{equation}
and from it, let us define the following dimensionless variables
\begin{eqnarray} 
Q\equiv\frac{H}{D} \,, & ~~ & \tilde{\Omega}_\rho\equiv
\frac{\kappa^2\rho}{3D^2}\,,  \label{ncvd} \\
\tilde{\Omega}_\Lambda\equiv
\frac{\Lambda}{3D^2}\,, & ~~ & \tilde{\Omega}_\lambda\equiv
\frac{1}{6\lambda}\frac{\kappa^2\rho^2}{D^2} \,. \nonumber
\end{eqnarray}
From these definitions we see that now the case $H=0$ is included.
Moreover, the Friedmann equation takes the following form
\begin{equation} 
\tilde{\Omega}_\rho+\tilde{\Omega}_\Lambda+\tilde{\Omega}_\lambda
=1 \,, \label{fri3}
\end{equation}
which, together with the fact that $-1 \leq Q \leq 1$ [see 
Equation~(\ref{ncvd})], implies that the state space defined by the
new variables is indeed compact.  Using the following new time derivative
\begin{equation}
 ' \equiv \frac{1}{D}\frac{d}{dt} \,, \label{tde2}
\end{equation}
the system of evolution equations for the variables $D$, $Q$,
$\tilde{\Omega}_\rho$, $\tilde{\Omega}_\Lambda$, and
$\tilde{\Omega}_\lambda$ is given by
\begin{eqnarray}
D' & = & -(1+qQ^2) Q D \,, \nonumber \\
Q' & = & -qQ^2(1-Q^2) \,, \label{qqp2} \\
\tilde{\Omega}'_\rho & = & \left[2(1+qQ^2)-3\gamma\right]
Q\tilde{\Omega}_\rho\,,
\label{orp2} \\
\tilde{\Omega}'_\Lambda & = & 2(1+qQ^2)Q\tilde{\Omega}_\Lambda\,,
\label{oap2} \\
\tilde{\Omega}'_\lambda & = & 2\left[1+qQ^2-3\gamma\right]Q
\tilde{\Omega}_\lambda \,, \label{olp2} 
\end{eqnarray}
where the deceleration parameter is now given by
\[ 1+qQ^2 = \frac{3\gamma}{2}(\tilde{\Omega}_\rho+2\tilde{\Omega}_\lambda)\,,
\]
The evolution equation for $D$ is not coupled to the rest, so we will 
not consider it for the dynamical study.  Thus, we study the 
dynamical system for the variables $\mb{\tilde{\Omega}}\equiv(Q,
\tilde{\Omega}_\rho,\tilde{\Omega}_\Lambda,\tilde{\Omega}_\lambda)$, 
determined by the equations~(\ref{qqp2}-\ref{olp2}). The complete set of 
critical points, their coordinates in the state space, 
i.e.~$\mb{\tilde{\Omega}}{}^\ast$, and their corresponding eigenvalues 
are given in the following table~\cite{Nota}
\begin{quasitable}
\begin{tabular}{ccc}
Model & Coordinates & Eigenvalues  \\ \tableline
$\mbox{F}_\epsilon$ & $(\epsilon,1,0,0)$ & $\epsilon(3\gamma-2,3\gamma,
3\gamma,-3\gamma)$ \\
$\mbox{dS}_\epsilon$ & $(\epsilon,0,1,0)$ & $-\epsilon(2,3\gamma,0,6\gamma)$ 
\\
$\mbox{E}~$ & $(0,\tilde{\Omega}^\ast_\rho,\tilde{\Omega}^\ast_\Lambda,
\tilde{\Omega}^\ast_\lambda)$ & $ (0,\sqrt{\phi},0,-\sqrt{\phi})$ \\
$\mbox{m}_\epsilon$ & $(\epsilon,0,0,1)$ & $2\epsilon(3\gamma-1,
\textstyle{3\gamma\over2},3\gamma,3\gamma)$
\end{tabular}
\end{quasitable}
Where $\tilde{\Omega}^\ast_\rho$, $\tilde{\Omega}^\ast_\Lambda$ and 
$\tilde{\Omega}^\ast_\lambda$ are constants satisfying~(\ref{fri3}) and 
the relations 
\begin{equation} 
\tilde{\Omega}^\ast_\rho \ = \ 
2\left(\frac{1}{3\gamma}-\tilde{\Omega}^\ast_\lambda\right) \,, ~~~~
\tilde{\Omega}^\ast_\Lambda \ = \  1 - 
\frac{2}{3\gamma}+\tilde{\Omega}^\ast_\lambda \,. \label{pcei}
\end{equation}
Here, $\mbox{E}$ represents a set of infinite saddle points whose line
element is that of the Einstein universe ($k=1$ and $H=0$).  The eigenvalues
of these points are determined by $\phi$, which in terms of 
$\mb{\tilde{\Omega}}{}^\ast$ and $\gamma$ is given by
\[ \phi  \ = \  \frac{3\gamma}{2}\left[ \left(3\gamma-2\right)
\tilde{\Omega}^\ast_\rho+4\left(3\gamma-1\right)
\tilde{\Omega}^\ast_\lambda\right] \,. \]
One can check, using~(\ref{fri3}) and~(\ref{pcei}), that $\phi$ is always
positive.  The dynamical character of all the equilibrium points is given 
in the table below.  As we can see from the previous tables, it depends 
on the equation of state (on the parameter $\gamma$) and on the expanding 
($H>0$ $\Leftrightarrow$ $\epsilon=1$) or contracting character ($H<0$
$\Leftrightarrow$ $\epsilon=-1$) of the point:
\begin{quasitable}
\begin{tabular}{cccc}
Model  & \multicolumn{3}{c}{Dynamical character} \\
\mbox{} &  $0<\gamma < \textstyle{1\over3}$ & $\gamma=\textstyle{1\over3}$ & 
$\gamma > \textstyle{1\over3}$  \\ \tableline 
$\mbox{F}_{\pm}$ & saddle & saddle & saddle  \\ 
$\mbox{M}_+$  & repeller & repeller & saddle   \\
$\mbox{M}_-$  & attractor & attractor & saddle  \\
$\mbox{dS}_+$ & attractor & attractor & attractor  \\
$\mbox{dS}_-$ & repeller & repeller & repeller \\
$\mbox{E}$~   & \mbox{---}  & saddle & saddle  \\ 
$\mbox{m}_+$  & saddle & repeller & repeller  \\
$\mbox{m}_-$  & saddle & attractor & attractor 
\end{tabular}
\end{quasitable}
At this point, we can observe some differences with the general relativistic
case~\cite{GolEll:1999}.  First, the Einstein Universe (E) appears to be a 
critical point for $\gamma\geq \textstyle{1\over3}$, in contrast with
general relativity, where it appears for $\gamma\geq\textstyle{2\over3}$.
On the other hand, as we will discuss in detail later, the dynamical 
character of some of the points changes with respect to general 
relativity.  For instance, in the brane-world scenario the expanding 
and contracting flat FLRW models ($\mbox{F}_+$ and $\mbox{F}_-$ 
respectively) are no longer repeller and attractor, respectively, for 
$\gamma>\textstyle{2\over3}$. They are now saddle points.

Another important difference is that now we have additional critical points,
namely, $\mbox{m}_+$ and $\mbox{m}_-$.  Let us analyze in detail the 
dynamics represented by these models.  First of all, we have to
point out that their characterization presents an extra difficulty
with respect to the other models.  Their coordinates in the state
space are $\mb{\Omega^\ast}=(0,0,0,1)$ and $\mb{\tilde{\Omega}{}^\ast}=
(\epsilon,0,0,1)$, i.e. the contributions of the ordinary
matter term ($\Omega_\rho$), the spatial curvature ($\Omega_k$) and the 
cosmological constant ($\Omega_\Lambda$) are negligible. Therefore,
we have at the same time $\kappa^2\rho H^{-2}\rightarrow 0$ and 
$(6\lambda)^{-1}\kappa^2\rho^2H^{-2}\rightarrow 1$,
hence their characterization must involve a limiting process. 
In order to understand the dynamics let us consider the simplified 
situation $\Lambda={}^3R=0$, in which the Friedmann equation~(\ref{frie}) 
can be solved to give
\begin{equation}
a(t) = \left(t-t_{BB}\right)^{\frac{1}{3\gamma}}
\left(t+t_{BB}\right)^{\frac{1}{3\gamma}}\,, \label{adis}
\end{equation}
where the constant $t_{BB}$ is the Big-Bang time
\[ t_{BB}\equiv\sqrt{\frac{2}{3\gamma^2\kappa^2\lambda}} = 
\frac{1}{3\gamma}\left(\frac{\kappa_{(5)}}{\kappa}\right)^2 \,. \]
In the state-space diagrams shown in the Figures~\ref{esp1}-\ref{esp5} 
below, this situation corresponds to models in the line joining 
$\mbox{m}_+$ and $\mbox{F}_+$.  
From~(\ref{adis}), we deduce that for late times, $t\gg t_{BB}$,
the scale factor behaves as $a(t)\sim t^{\frac{2}{3\gamma}}\,,$ and
therefore the solution approaches the flat FLRW model ($\mbox{F}_+$),
hence we have a general relativistic behaviour.  However, the new
interesting behaviour appears when we approach the initial 
singularity ($t\rightarrow t_{BB}$) or, in other words, at very high 
energies ($\rho \gg \lambda$), where we have 
$a(t)\sim (t-t_{BB})^{\frac{1}{3\gamma}}\,.$ From the point of view of 
Einstein's equations~(\ref{mefe}), in such a situation
the term involving the four-dimensional constant, $\kappa\,,$ becomes 
negligible with respect to the term involving the five-dimensional one, 
$\kappa_{(5)}\,.$ 
We recover general relativity in the limit $t_{BB}\rightarrow 0$, which 
is the opposite situation.  From this discussion we realize that the 
limiting process leading to the critical points $\mbox{m}_\pm$ is
\begin{equation}
\kappa ~ \rightarrow ~ 0 ~~~~~ \left(\frac{\kappa^2}{6\lambda}
\rightarrow \frac{1}{\kappa^4_{(5)}}\right)\,. \label{thel}
\end{equation}
Then, we find that the points $\mbox{m}_\pm$ are models whose
scale factor is given by
\begin{equation} 
a(t) = t^{\frac{1}{3\gamma}}\,. \label{brex}
\end{equation}
This is the Bin\'etruy-Deffayet-Langlois (BDL) solution~\cite{BinDefLan:2000} 
(see also~\cite{BinDefEllLan:2000,FlaTyeWas:2000b})
As we have already mentioned, these models describe
the dynamics near the singularities.  That is, the early universe 
behaviour, near the initial Big-Bang singularity and also,
for recollapsing models (for which we must have ${}^3R>0$),
the dynamical behaviour when we approach the Big-Crunch singularity.  
In both cases the dynamics changes with respect to general relativity.

With the information we have obtained about the critical points of
the dynamical systems for $\mb{\Omega}$ and $\mb{\tilde{\Omega}}$, we 
can apply the well-known techniques used in dynamical 
systems~\cite{DSTH} to obtain the structure of the 
state space, which provides, in a visual way, the complete 
information on the evolution of our system (a perfect-fluid FLRW model 
in the brane-world scenario) once the initial conditions are given.  
In the same way as the dynamical character of the critical points 
depend on the equation of state, or equivalently,
on the parameter $\gamma$, so will do the state space.  In fact, we
have found that there are values of $\gamma$ for which 
{\em bifurcations}, that is, topological changes in the state space
(see~\cite{DSTH} for details), appear.  Specifically, 
these values are $\gamma_B=0,\textstyle{1\over3},\textstyle{2\over3}$
(in general relativity we only have bifurcations at $\gamma=0,
\textstyle{2\over3})$.  As we will see in the discussion of each
particular case, for $\gamma=\textstyle{1\over3}$ and
$\gamma=\textstyle{2\over3}$, we have lines with an infinite number of
critical points, for which we get one vanishing eigenvalue, as is expected
in those cases~\cite{DSTH}.

Let us begin with the $\gamma=0$ case.  We have not drawn the state
space because it is quite simple.  Equation~(\ref{emce})
implies that the energy density is constant.  Then, we can solve the 
Friedmann equation~(\ref{frie}) and we find that $a(t)$ is given by
\begin{equation}
a(t) = \left\{ \begin{array}{ll}
\sqrt{\frac{3}{\tilde{\Lambda}}} \cosh\left[\epsilon
\sqrt{\frac{\tilde{\Lambda}}{3}}(t-t_i)\right]~~ & \mbox{for $k=1$}\,, \\
\mbox{e}^{\epsilon\sqrt{\frac{\tilde{\Lambda}}{3}}(t-t_i)} 
& \mbox{for $k=0$}\,, \\
\sqrt{\frac{3}{\tilde{\Lambda}}} \sinh\left[\epsilon
\sqrt{\frac{\tilde{\Lambda}}{3}}(t-t_i)\right] & \mbox{for $k=-1$}\,, 
\end{array} \right. \label{dscl}
\end{equation}
where $t_i$ is a constant, $\epsilon=\mbox{sgn}(H)$, and 
$\tilde{\Lambda}$ is a modified cosmological constant given by
\begin{equation}
\tilde{\Lambda} = \Lambda +\kappa^2\rho\left(1 +
\frac{\rho}{2\lambda}\right) \,. \label{mcco}
\end{equation}
For $\tilde{\Lambda}\neq 0$, all the models belong to the de Sitter class, 
whereas in the limit $\tilde{\Lambda}\rightarrow 0$ ($\Lambda=\rho=0$) 
we find the Minkowski ($k=0$) and Milne ($k=-1$) spacetimes. 
The dynamics (of expanding models, $\epsilon=1$) is reduced to the fact 
that the model $k=0$ is the future attractor, and the Milne universe 
is a repeller. 

For the other cases ($\gamma\neq 0$), the whole state space is constructed
by matching the state space corresponding to the dynamical 
systems~(\ref{orp1}-\ref{oap1}) and~(\ref{qqp2}-\ref{olp2}). It
consists of three pieces, the diagram shown in Figure~\ref{esp1}(b) 
on the right, which corresponds to the case $\epsilon=1$ 
in~(\ref{orp1}-\ref{oap1}), the diagram in Figure~\ref{esp1}(a) in
the middle, and on the left the diagram corresponding to the
case $\epsilon=-1$ in~(\ref{orp1}-\ref{oap1}), which has not
been included here because it can be obtained from the 
Figure~\ref{esp1}(b) just by reversing the direction of the arrows
and replacing the subscript ``$+$" by ``$-$".  In order to follow
the evolution, we have specified the quantities represented in the
different axes.  Notice that the state space
is compact, with the boundaries given by the planes
$\Omega_\Lambda=\tilde{\Omega}_\Lambda=0$, 
$\Omega_\lambda=\tilde{\Omega}_\lambda=0$
and the vacuum models $\Omega_\rho=\tilde{\Omega}_\rho=0$.

We have drawn only the trajectories on 
the planes, but the trajectory of any point in the state space
outside these planes can be deduced qualitatively following the behaviour 
shown in them.   As is obvious, the general relativistic state 
space corresponds to the plane $\Omega_\lambda=0\,,$ which is an
{\em invariant  submanifold}\footnote{State space trajectories starting
in an invariant submanifold will never leave it.} of the state space.
Therefore, the aim of this work is to study what happens when we take 
initial conditions outside of this plane.  The other invariant 
submanifolds are: the vacuum boundary $\Omega_\rho=0$, the flat geometry 
submanifold $\Omega_k=0$, and the $\Omega_\Lambda=0$ submanifold.

Keeping this preamble in mind, let us analyze the different cases according
to $\gamma\,.$ For $\gamma\in(0,\textstyle{1\over3})$ and ${}^3R\leq 0$,
Milne is a repeller, as in the general-relativistic case, and the expanding
de Sitter model is the future attractor for all the initial conditions
excepting the plane $\Omega_\Lambda=0\,,$ for which the attractor is the flat
FLRW model. For ${}^3R>0\,,$ $\mbox{dS}_+$ plays the same role.  In the 
plane $\Omega_\Lambda=0$, collapsing FLRW models evolve towards 
the expanding flat FLRW model ($\mbox{F}_+$), with the effect of 
the extra dimension being maximum when $H=0$ ($\Leftrightarrow Q=0$).
In conclusion, the dynamics in this case is essentially the same as
in general relativity.

\begin{figure*}
\begin{center}
\includegraphics[height=3in,width=3in, bbllx=55, bblly=200, bburx=540, 
bbury=680]{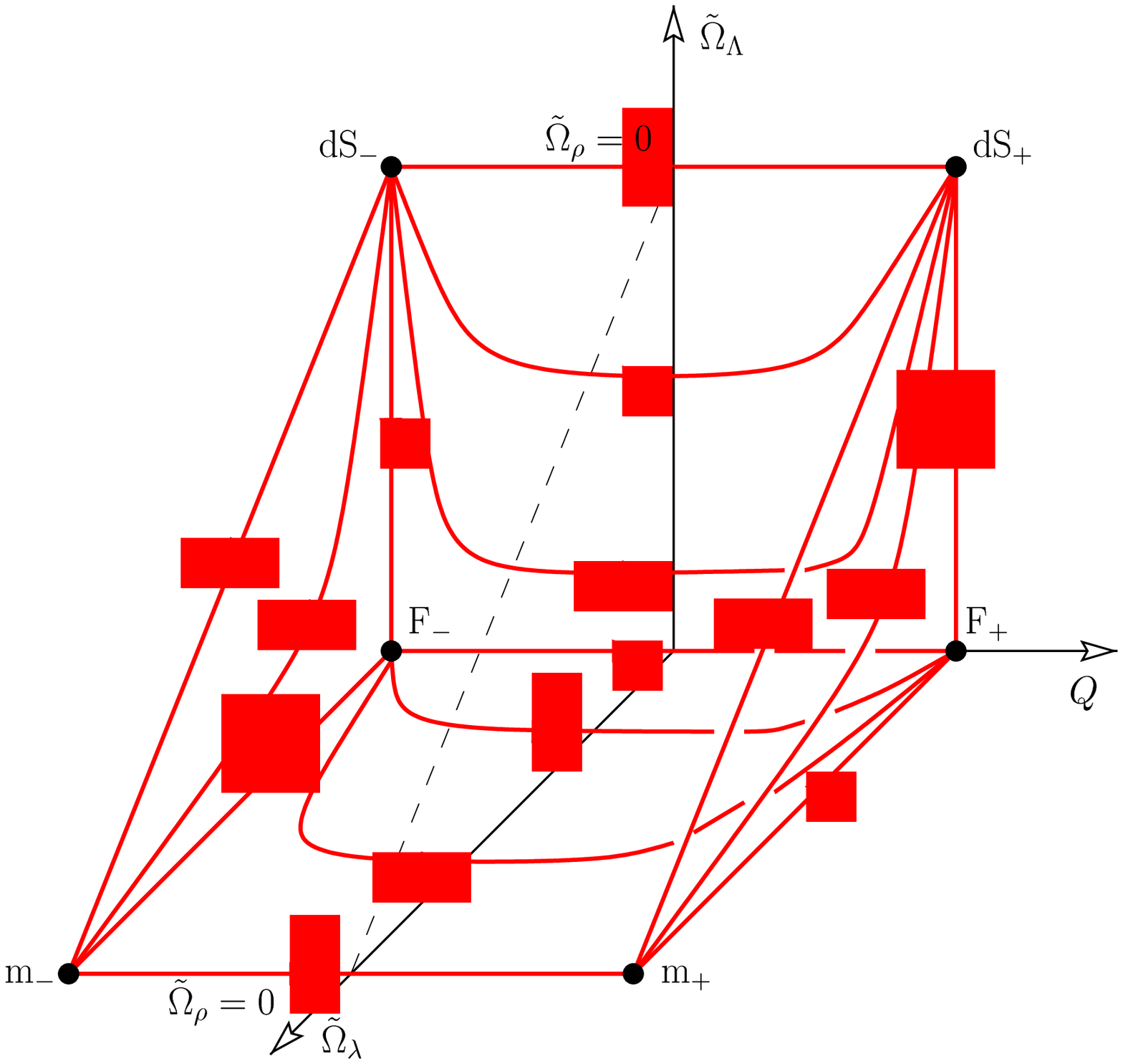}
\includegraphics[height=3in,width=3in, bbllx=55, bblly=200, bburx=540, 
bbury=680]{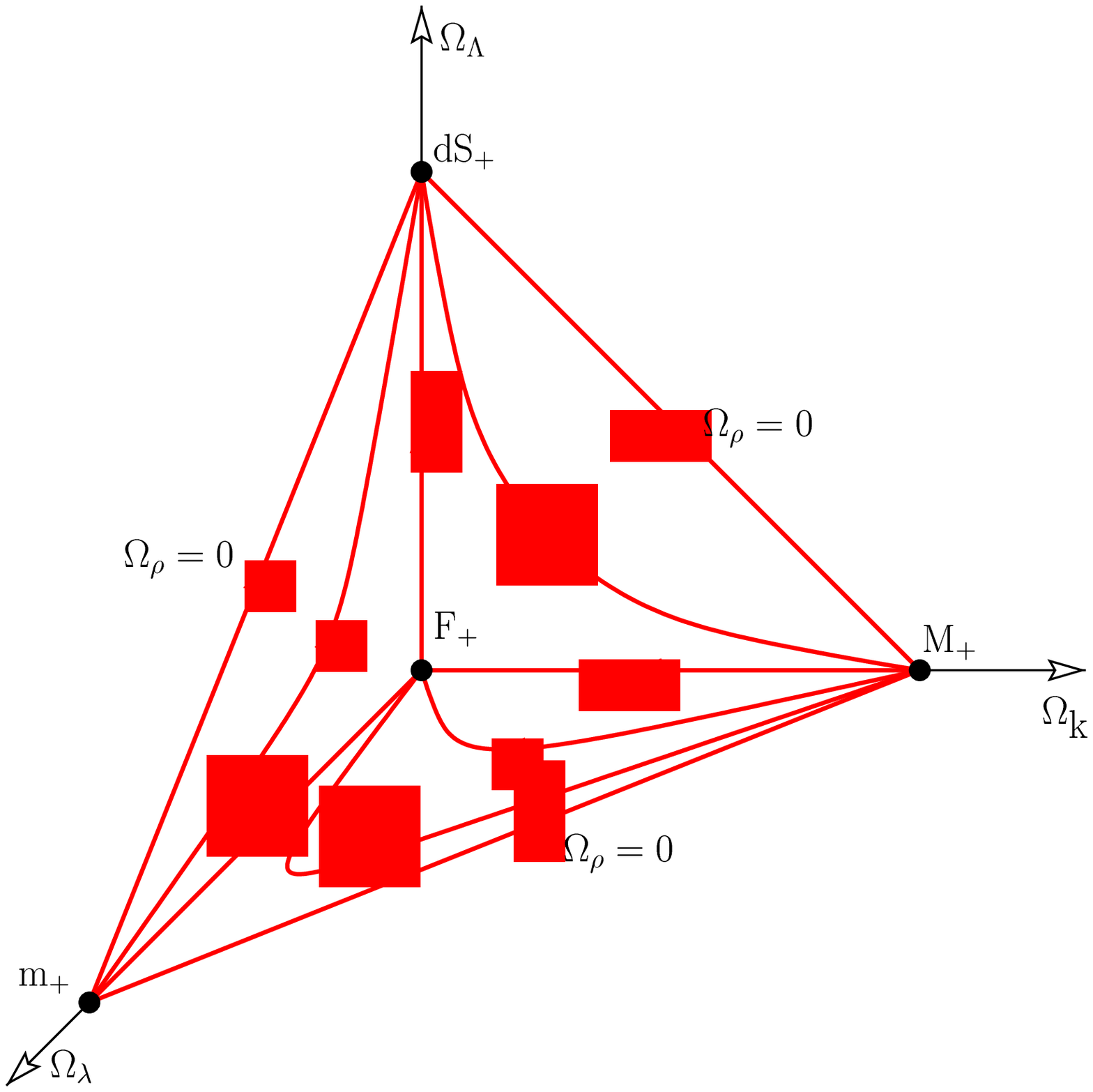}
\caption{State space for the FLRW models with 
$\gamma\in(0,\textstyle{1\over3})$ and (a) 
non-negative spatial curvature, ${}^3 R\geq 0$ (on the left) and (b) 
non-positive spatial curvature, ${}^3 R\leq 0$ (on the right).
Replacing $\Omega_k$ by $\Omega_\sigma$ and $\mbox{M}_\epsilon$ by 
$\mbox{K}_\epsilon$ the drawing on the right is also the state space 
for Bianchi I models with $\gamma\in(0,1)\,.$
The planes $\tilde{\Omega}_\Lambda - Q$ and $\Omega_\Lambda - \Omega_k$ 
correspond to the state space of general relativity. 
The critical points $\mbox{F}_\epsilon$, $\mbox{M}_\epsilon$, 
$\mbox{dS}_\epsilon$, $\mbox{E}$, $\mbox{m}_\epsilon$ and 
$\mbox{K}_\epsilon$ describe the flat 
FLRW model, the Milne universe, the de Sitter model, the Einstein universe, 
the non-general-relativistic BDL models and the Kasner spacetimes 
respectively. 
$\epsilon$ is the sign of the Hubble function, differentiating between
expanding and collapsing models. The planes joining the points 
$\mbox{dS}_\epsilon$, $\mbox{M}_\epsilon$ and $\mbox{m}_\epsilon$
represent vacuum solutions ($\Omega_\rho=\tilde{\Omega}_\rho=0$).
Only trajectories on the invariant planes, which outline the whole 
dynamics, are drawn ($\Omega_k=0$, 
$\Omega_\Lambda=\tilde{\Omega}_\Lambda=0$, and $\Omega_\lambda=
\tilde{\Omega}_\lambda=0$).}
\label{esp1}
\end{center}
\end{figure*}

The next case, $\gamma=\textstyle{1\over3}\,,$ constitutes a bifurcation.
The topology of the state space changes [see Figures~\ref{esp2}(a)
and~\ref{esp2}(b)] due to the fact that we have
now a line of vacuum critical points.  This line extends to the 
three parts of the whole state space.  In the ${}^3R\geq 0$ sector,
Figure~\ref{esp2}(a), all these critical points, excepting the points
$\mbox{m}_\pm$ and $\mbox{E}$, are not included in the previous tables.
Their coordinates are $\mb{\tilde{\Omega}{}^\ast}=(Q^\ast,0,0,1)\,,$ 
$|Q^\ast|<1$, and hence they do not appear in general relativity.   
In order to see to what particular models they corresponds we need to 
consider the limit~(\ref{thel}) since they have 
$\tilde{\Omega}{}^\ast_\lambda=1$.  Then, solving the Friedmann 
equation~(\ref{frie}), we find they are positively curved FLRW
models with dynamics described by $a(t)\sim t\,.$ The particular case
$Q^\ast=0$ corresponds to the Einstein universe.
In the ${}^3R\leq 0$ sector, Figure~\ref{esp2}(b), the critical points, 
excepting points $\mbox{m}_\pm$ and $\mbox{M}_\pm$, are also not in the 
tables above and they are non-general-relativistic in nature.
Their coordinates are 
$\mb{\Omega^\ast}=(0,\Omega_k^\ast,0,\Omega_\lambda^\ast)$ with
$\Omega_k^\ast+\Omega_\lambda^\ast=1$.  Then, using the
limiting procedure~(\ref{thel}), we get the same time
dependence: $a(t)\sim t\,.$ These points are also non-general-relativistic.

\begin{figure*}
\begin{center}
\includegraphics[height=3in,width=3in, bbllx=55, bblly=200, bburx=540, 
bbury=680]{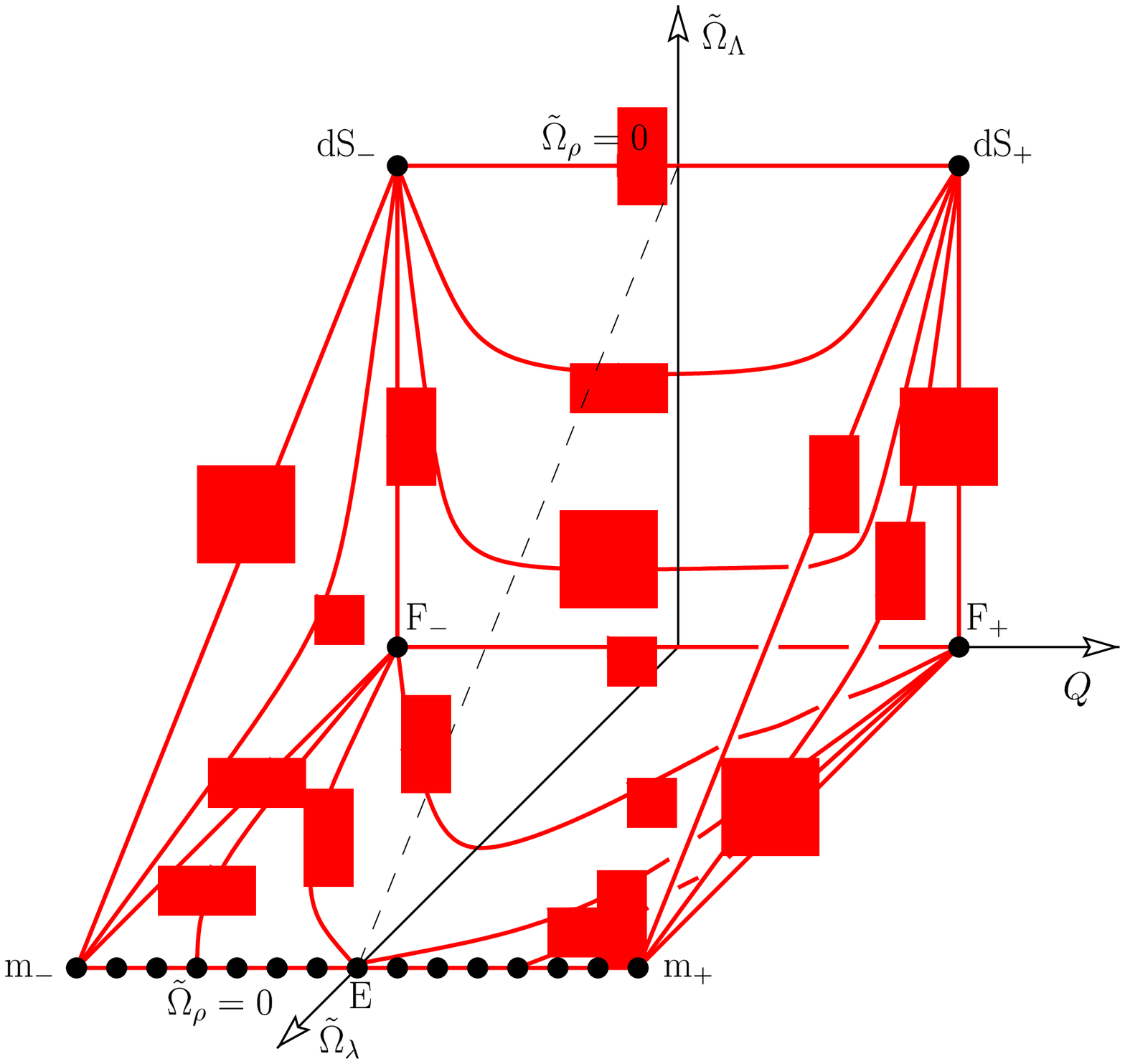}
\includegraphics[height=3in,width=3in, bbllx=55, bblly=200, bburx=540, 
bbury=680]{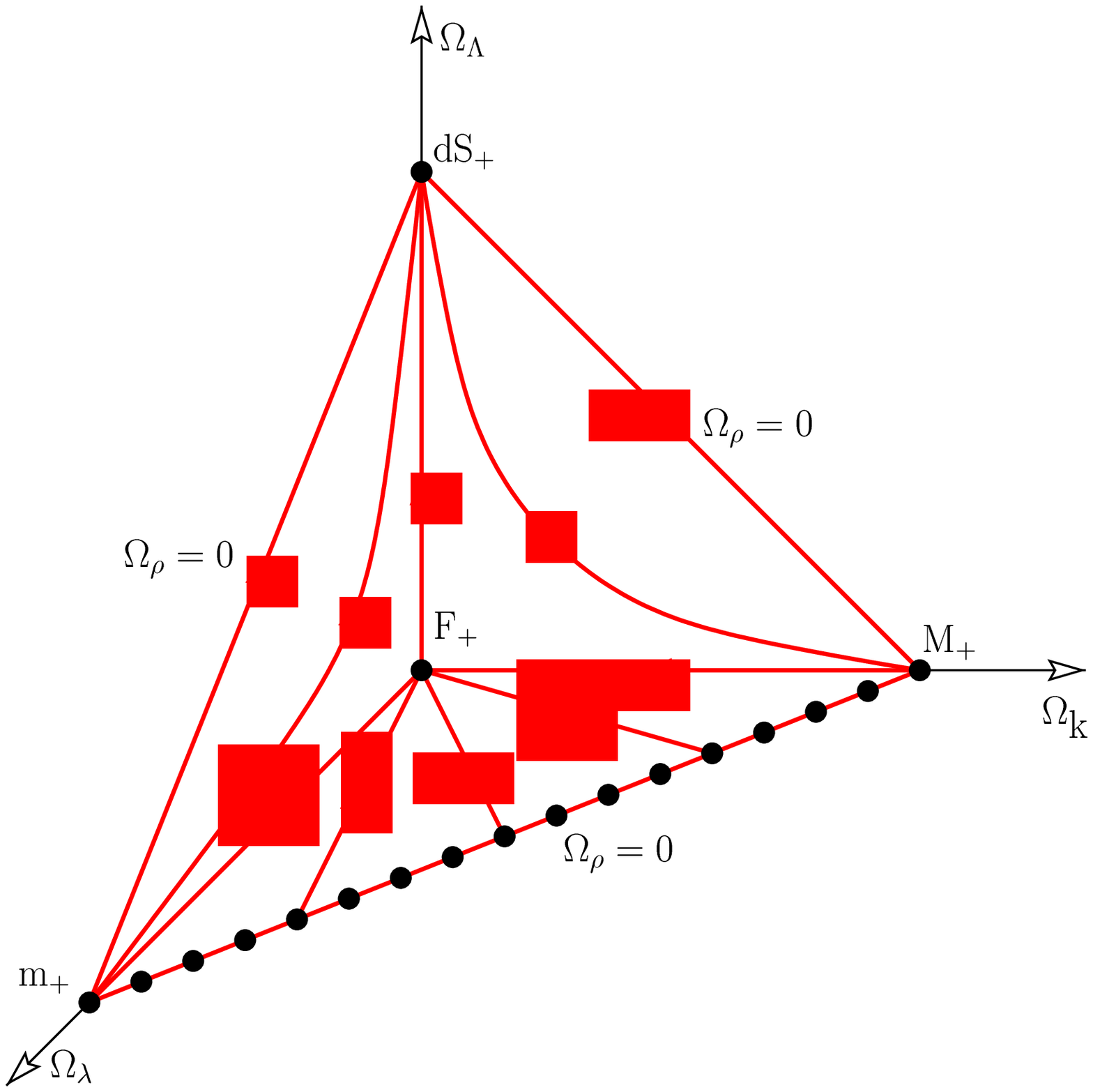}
\caption{State space for the FLRW models with $\gamma=\textstyle{1\over3}$ 
(a bifurcation) and (a) non-negative 
spatial curvature, ${}^3 R\geq 0$ (on the left) and (b) non-positive 
spatial curvature, ${}^3 R\leq 0$ (on the right). The drawing on the right
is also the state space for Bianchi models with $\gamma=1$. See the caption
of Figure~\ref{esp1} for more details.}\label{esp2}
\end{center}
\end{figure*}

The next step is to study the state space in the interval 
$\gamma\in(\textstyle{1\over3},\textstyle{2\over3})$, which is now
described by the diagrams shown in Figure~\ref{esp3}.  
We can find some changes with respect to the situation in the previous 
cases.  First, we have an infinite number of critical points 
corresponding to the Einstein universe, which are arranged in a line 
determined by
equations~(\ref{pcei}).  In the ${}^3R\geq 0$ sector we can see that
$\mbox{dS}_+$ and $\mbox{m}_-$ are attractors of the evolution.
Then, this sector of the state space is divided into two regions.
The first one consists of those points which will evolve to the
de Sitter model ($\mbox{dS}_+$), which corresponds to the whole
${}^3R\geq 0$ sector in the case $\gamma\in(0,\textstyle{1\over3})$.  The 
second region is determined by the points which evolve towards the 
BDL model ($\mbox{m}_-$) which does not contain any general relativistic 
point.  These trajectories correspond to models collapsing in the future, 
that is, evolving towards a Big Crunch singularity, where the dynamics 
is given by~(\ref{brex}).  It is worth noting that in general relativity 
recollapsing models only occur for $\gamma > \textstyle{2\over3}\,.$
These two regions are separated by the surface generated by the trajectories
that start from or arrive to the
set of critical points representing the Einstein universe ($\mbox{E}$),
which are saddle points.  In the ${}^3R\leq 0$ sector the situation is
simpler. For a vanishing cosmological constant ($\Omega_\Lambda=0$)
the future attractor are the flat FLRW models ($\mbox{F}_+$) whereas
in the case with a cosmological constant it is the de Sitter model
($\mbox{dS}_+$).

\begin{figure*}
\begin{center}
\includegraphics[height=3in,width=3in, bbllx=55, bblly=200, bburx=540, 
bbury=680]{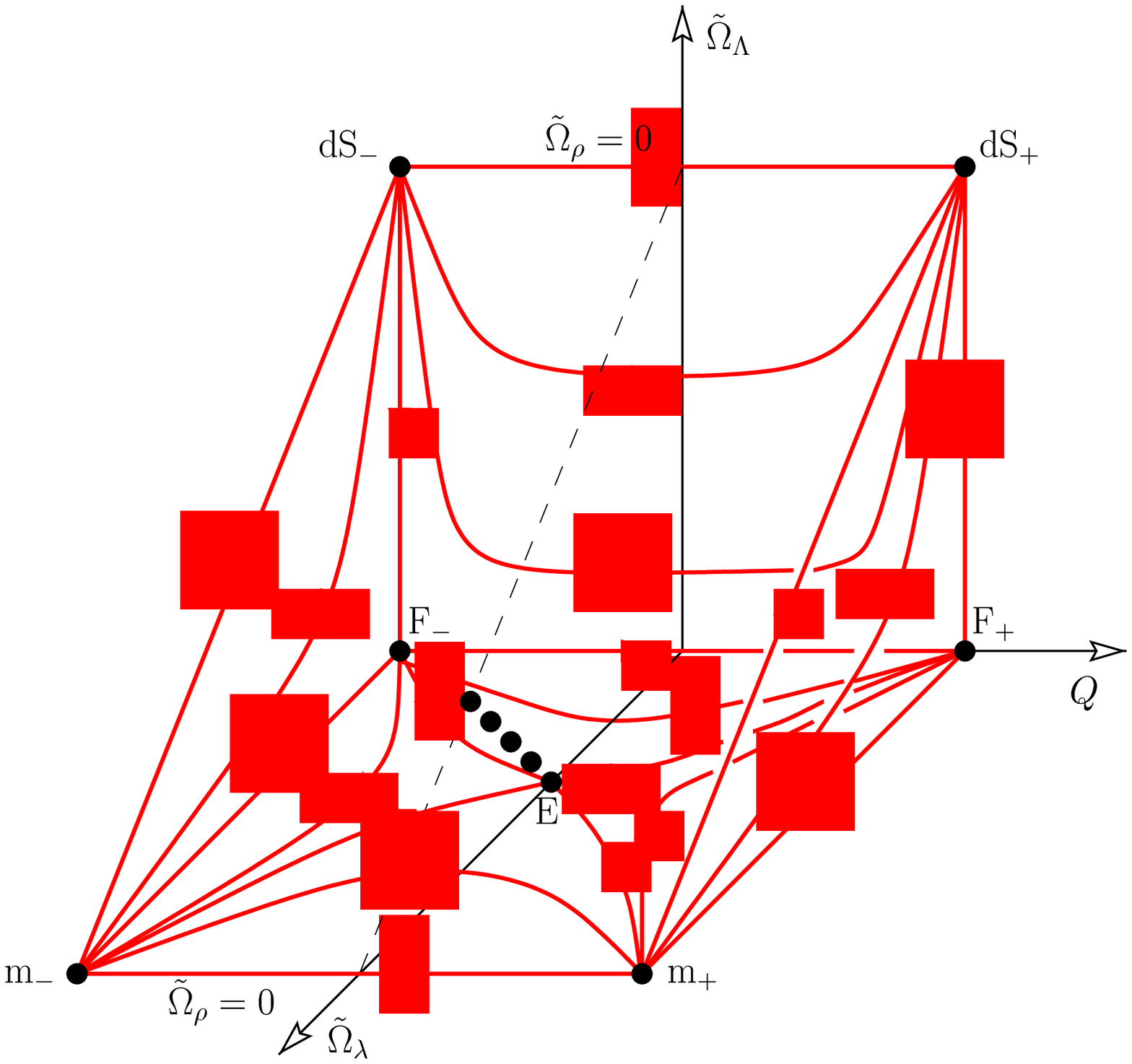}
\includegraphics[height=3in,width=3in, bbllx=55, bblly=200, bburx=540, 
bbury=680]{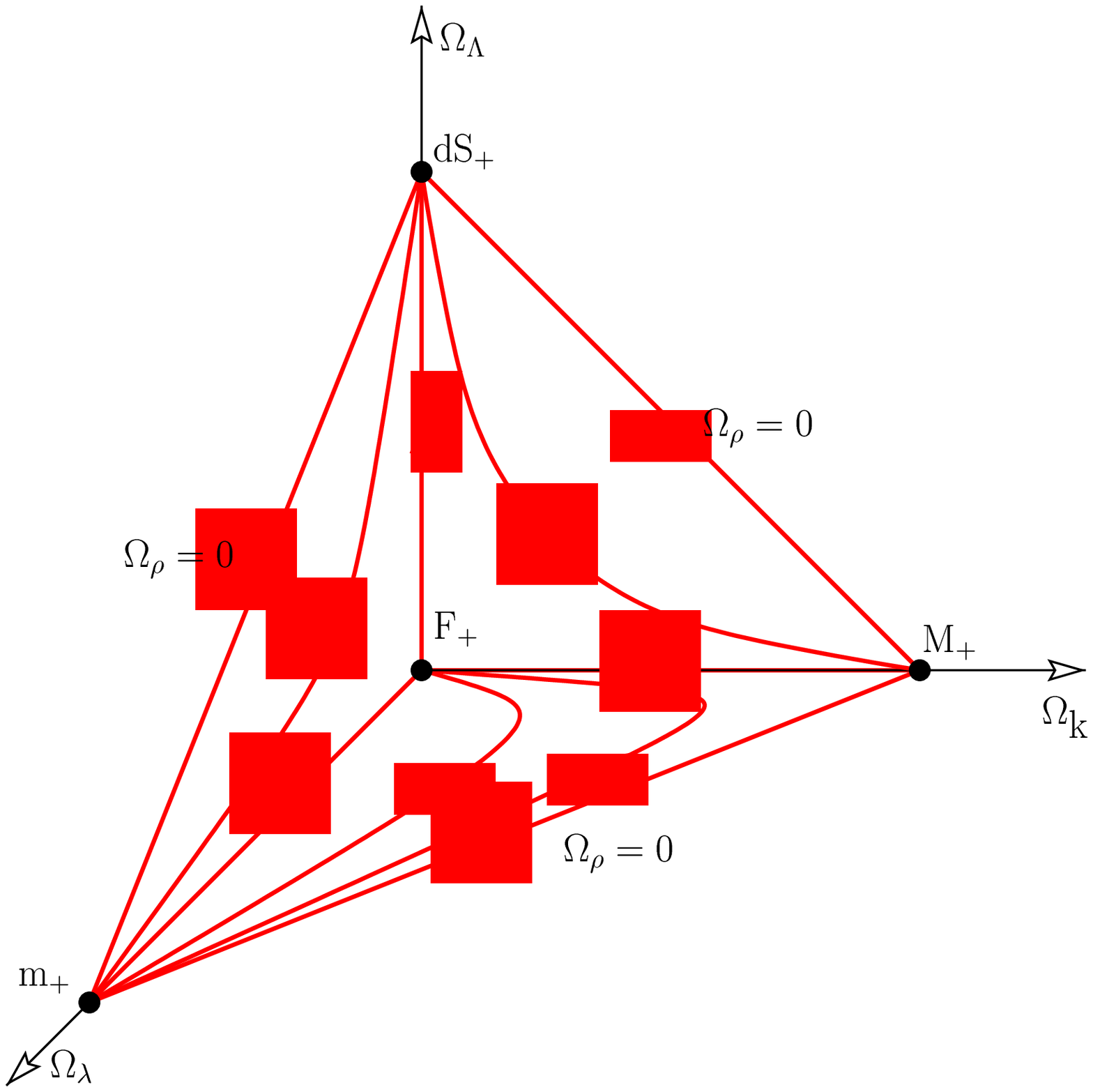}
\caption{State space for the FLRW models with 
$\gamma\in (\textstyle{1\over3},\textstyle{2\over3})$ 
and (a) non-negative spatial curvature, ${}^3 R\geq 0$ (on the left) 
and (b) non-positive spatial curvature, ${}^3 R\leq 0$ (on the right). 
The drawing on the 
right is also the state space for Bianchi models with $\gamma\in(1,2)$. 
See the caption of Figure~\ref{esp1} for more details.}\label{esp3}
\end{center}
\end{figure*}

In $\gamma=\textstyle{2\over3}$ we have another bifurcation motivated by the 
appearance of two lines of infinite critical points which join at the general
relativistic Einstein universe, given by $\mb{\tilde{\Omega}{}^\ast}=
(0,1,0,0)$ (see Figure~\ref{esp4}).
One of the lines is composed by Einstein universe points whose
state space coordinates satisfy~(\ref{pcei}).  The other line
corresponds to general relativistic models (which were not shown 
in~\cite{GolEll:1999}), and it occupies the three regions.  Their points are
characterized by $\Lambda=\lambda^{-1}=0\,,$ and their scale factor grows
linearly with time ($H\neq 0$), $a(t)=Ct$.  They are
perfect-fluid models with equation of state $\rho+3p=0$ and energy
density given by
\[ \kappa^2\rho = \frac{3(C^2+k)}{C^2t^2} \,. \]
The case $k=-1$ (${}^3R<0$) and $C=1$ corresponds to the Milne universe.

\begin{figure*}
\begin{center}
\includegraphics[height=3in,width=3in, bbllx=55, bblly=200, bburx=540, 
bbury=680]{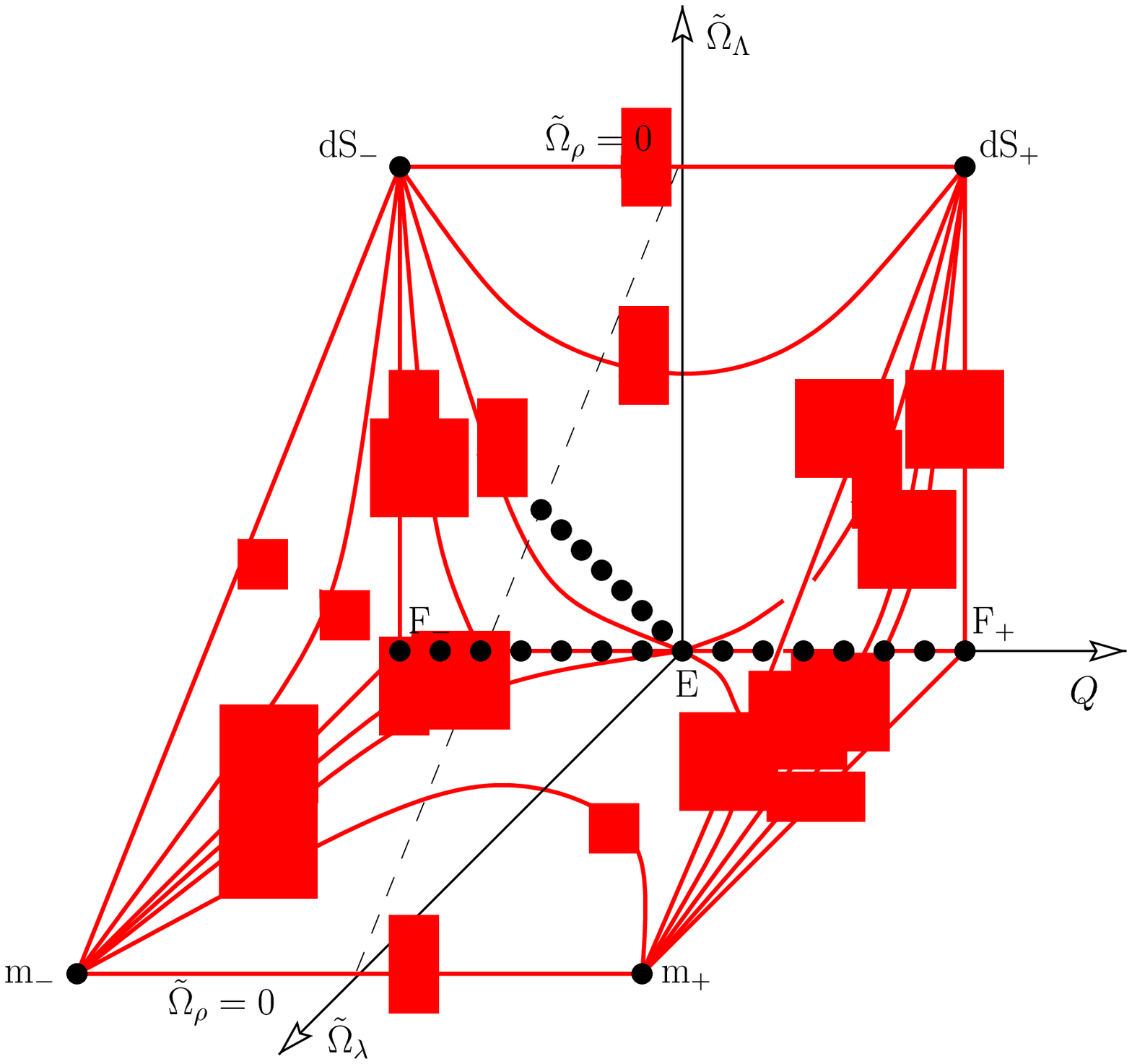}
\includegraphics[height=3in,width=3in, bbllx=55, bblly=200, bburx=540, 
bbury=680]{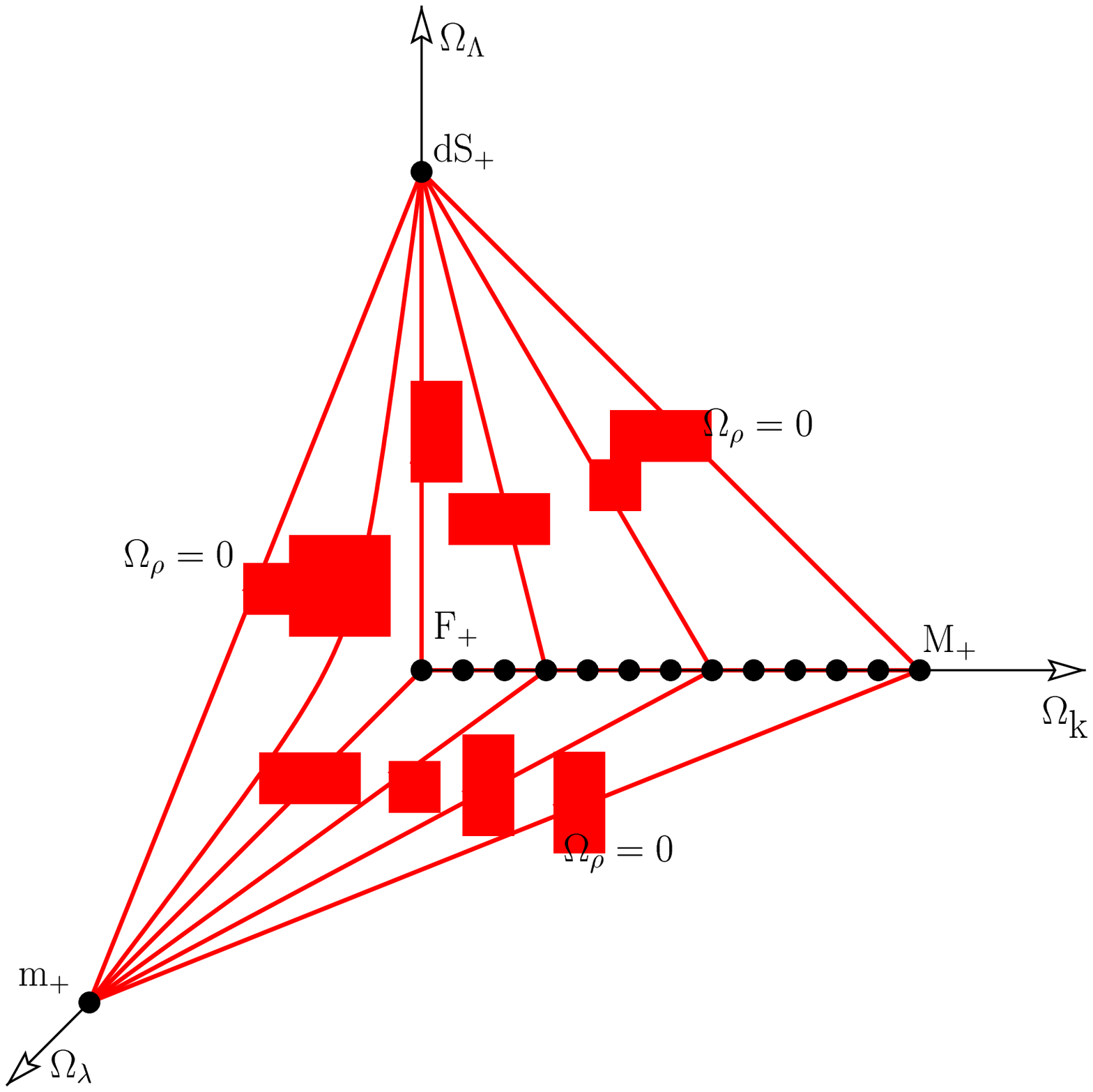}
\caption{State space for the FLRW models with $\gamma=\textstyle{2\over3}$ 
(a bifurcation) and (a) non-negative 
spatial curvature, ${}^3 R\geq 0$ (on the left) and (b) non-positive spatial 
curvature, ${}^3 R\leq 0$ (on the right). The drawing on the 
right is also the state space for Bianchi models with $\gamma=2$. 
See the caption of Figure~\ref{esp1} for more details.}\label{esp4}
\end{center}
\end{figure*}

The last situation corresponds to the case $\gamma>\textstyle{2\over3}$,
described by the state space drawn in Figure~\ref{esp5}.  The situation 
in the ${}^3R\geq 0$ sector is now very similar to that showed in the
$\gamma\in(\textstyle{1\over3},\textstyle{2\over3})$ case, where two regions 
appeared according to whether the points evolve to the BDL or 
to the de Sitter model.  The region of points evolving to the BDL 
model is now bigger.  With regard to the ${}^3R\leq 0$ sector,
the situation has now changed:  for the models without cosmological
constant the attractor is the Milne universe ($\mbox{M}_+$), whereas
the flat FLRW models ($\mbox{F}_+$) are saddle points.  For a
non-vanishing cosmological constant, the de Sitter model ($\mbox{dS}_+$)
is again the attractor.

\begin{figure*}
\begin{center}
\includegraphics[height=3in,width=3in, bbllx=55, bblly=200, bburx=540, 
bbury=680]{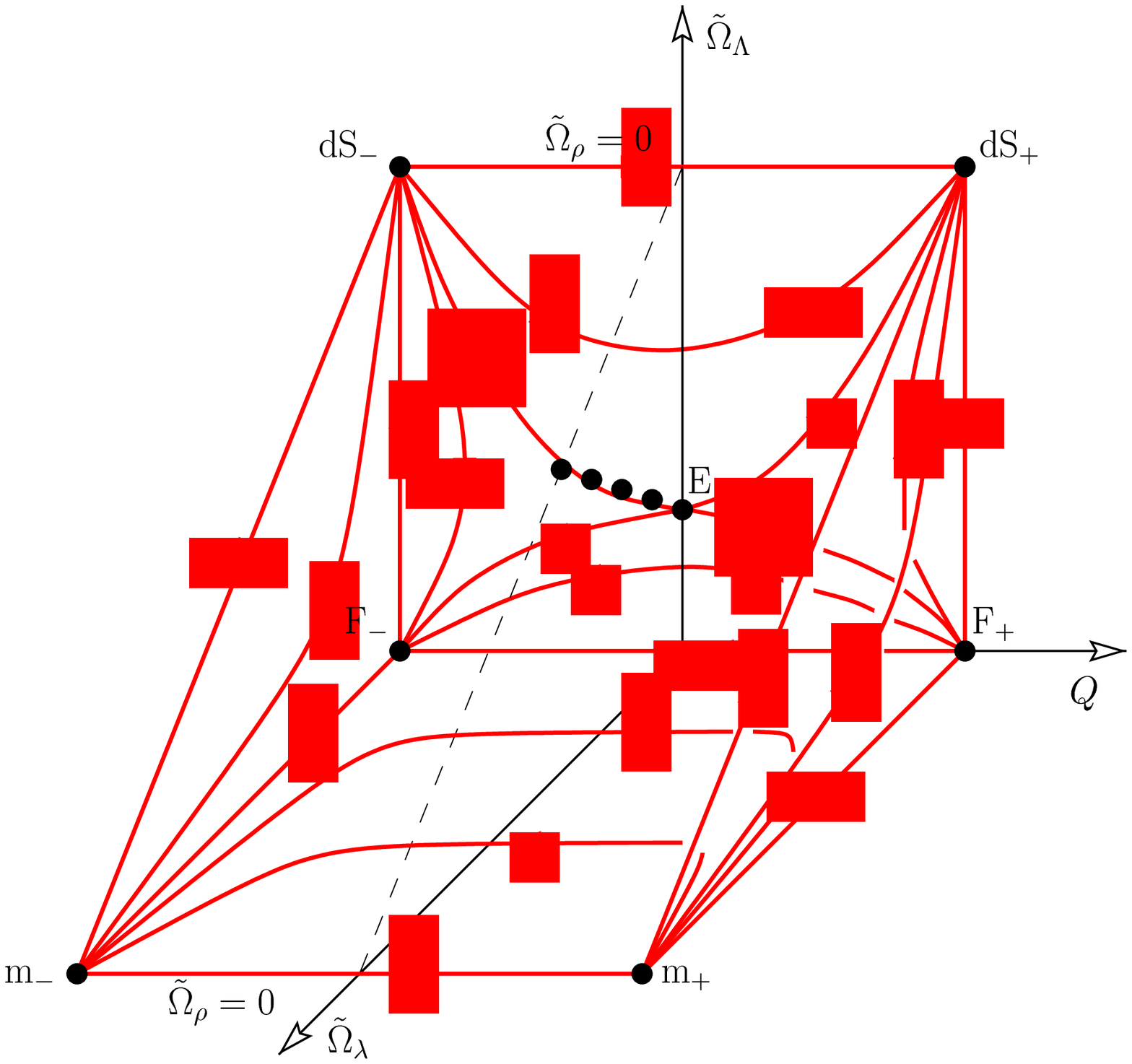}
\includegraphics[height=3in,width=3in, bbllx=55, bblly=200, bburx=540, 
bbury=680]{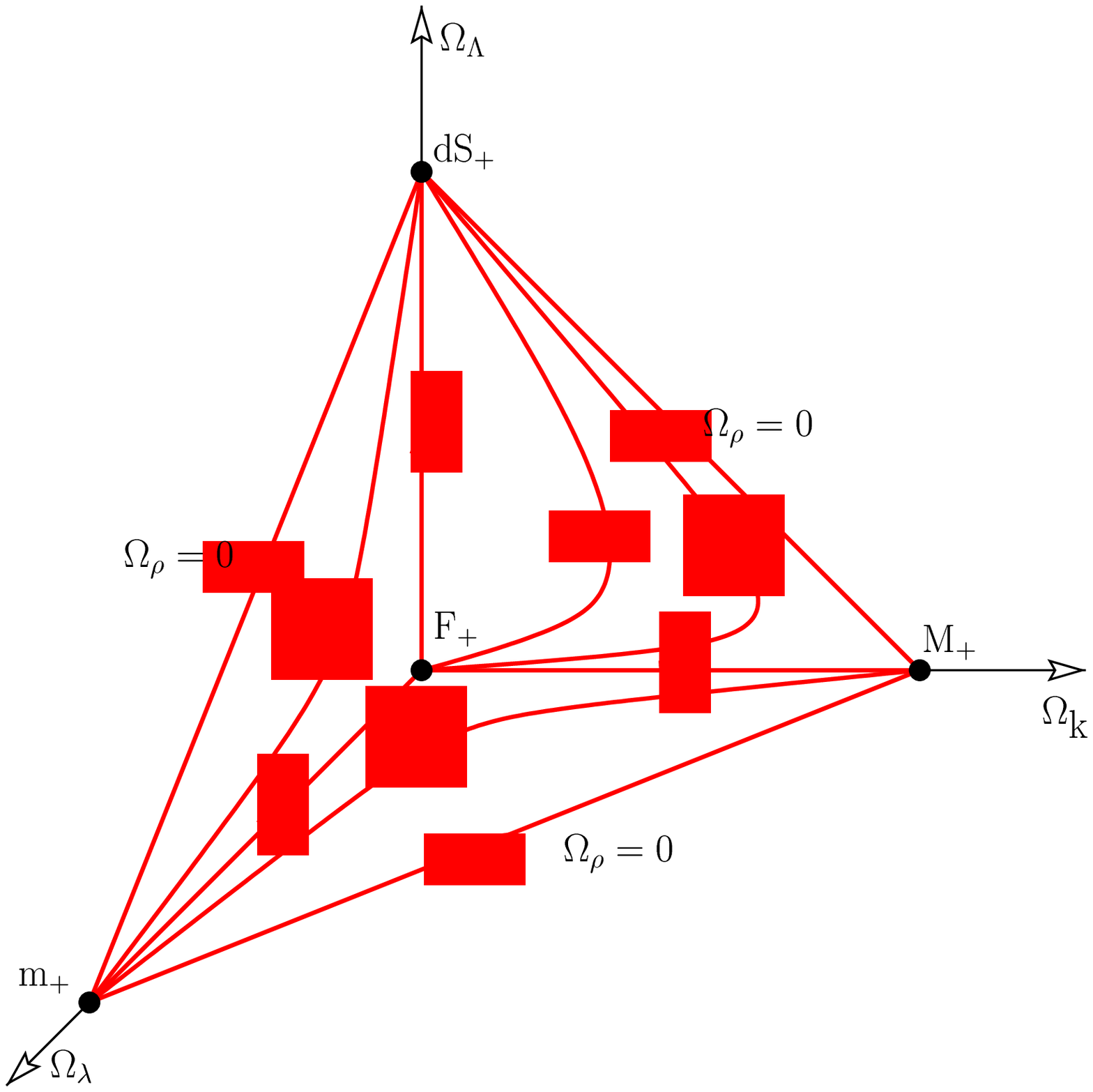}
\caption{State space for the FLRW models with $\gamma>\textstyle{2\over3}$ 
and (a) non-negative 
spatial curvature, ${}^3 R\geq 0$ (on the left) and (b) 
non-positive spatial curvature, ${}^3 R\leq 0$ (on the right).
See the caption of Figure~\ref{esp1} for more details.}\label{esp5}
\end{center}
\end{figure*}



\section{Dynamics of Bianchi models in the 
brane-world scenario}\label{sec3}
In this section we will
study the dynamics of some homogeneous but anisotropic cosmological models 
(Bianchi models) in the brane-world scenario.  In particular, we will 
consider the perfect-fluid Bianchi I and V homogeneous cosmological models 
in which the fluid velocity is non-tilted, which means that the 
hypersurfaces of homogeneity are orthogonal to the fluid flow.  Moreover, 
we will also consider a linear equation of state~(\ref{leoe}) for the 
perfect fluid.  We have considered these two particular classes of Bianchi 
models for simplicity and because they contain the flat and negatively 
curved FLRW models.

It is well-known that Bianchi models can be described by systems of ordinary 
differential equations, being the fluid proper time $t$ the only independent
variable that appears.  The form of the system of ordinary differential
equations depends on the parametrization of the models, i.e., on the 
variables we use to describe them.  Here, we will start using the point of 
view adopted by Ellis and MacCallum~\cite{EllMac:1969}, where 
they use an orthonormal tetrad, $\{\mb{u},\mb{e}_\alpha\}$ ($\alpha=
1,\ldots,3$), adapted to the fluid velocity 
\[ \mb{u}\cdot\mb{u}=-1\,,~~ \mb{u}\cdot\mb{e}_\alpha=0\,,~~
\mb{e}_\alpha\cdot\mb{e}_\beta = \delta_{\alpha\beta}\,. \]
Then, the dynamics can be described in terms of the following variables:
(i) The {\em spatial commutation functions}, 
$\gamma^\alpha{}_{\beta\delta}$, defined
by the commutation relations between the spatial basis vectors, 
$\left[\vec{\mbox{e}}_\beta,\vec{\mbox{e}}_\delta\right]=
\gamma^\alpha{}_{\beta\delta}\vec{\mbox{e}}_\alpha$
($\gamma^\alpha{}_{[\beta\delta]}=\gamma^\alpha{}_{\beta\delta}$).
Here, we will use the equivalent variables 
\[  a_\alpha\equiv\textstyle{1\over2}\gamma^\beta{}_{\alpha\beta} \,,
\hspace{4mm} n^{\alpha\beta}\equiv\textstyle{1\over2}
\varepsilon^{\delta\kappa(\alpha}\gamma^{\beta)}{}_{\delta\kappa}\,.\]
introduced by Sch\"ucking, Kundt and Behr (see~\cite{EllMac:1969} and 
references therein) to decompose $\gamma^\alpha{}_{\beta\delta}$ as
follows:
$\gamma^\alpha{}_{\beta\delta} = 2 a_{[\beta}\delta^\alpha{}_{\delta]}
+\varepsilon_{\beta\delta\kappa}n^{\alpha\kappa}\,.$ 
(ii) The {\em kinematical quantities}.  The Hubble function $H$ 
($\equiv \nabla_au^a/3$) and the 
components of the shear tensor $\sigma_{ab}$
\[ \sigma_{ab}\equiv h_a{}^ch_b{}^d\nabla_{(c}u_{d)}-Hh_{ab}\,, \] 
where $h_{ab}\equiv g_{ab}+u_au_b$ is the orthogonal projector to the 
fluid velocity $\mb{u}$. 
(iii) The {\em matter variables}.  In our case only the energy density 
$\rho$ and the isotropic pressure $p$, related by an equation of 
state~(\ref{leoe}).

In the case of Bianchi models the generalized Friedmann equation reads 
as follows
\begin{equation} 
H^2 = \frac{1}{3}\kappa^2\rho\left(1+\frac{\rho}{2\lambda}\right)
-\frac{1}{6}{}^3R+\frac{1}{3}\sigma^2+\frac{1}{3}\Lambda \,, \label{bfri}
\end{equation}
where $2\sigma^2\equiv\sigma^{ab}\sigma_{ab}$ and the spatial scalar
curvature has the following expression in terms of the spatial
commutation functions
\begin{equation} 
{}^3R=-6a^\alpha a_\alpha-n^{\alpha\beta}n_{\alpha\beta}+
\textstyle{1\over2}\left(n^\alpha{}_\alpha\right)^2  \,. \label{tcur}
\end{equation}
On the other hand, from the Einstein equations~(\ref{mefe}) we have a
constraint on our variables
\begin{equation} 
3\sigma_{\alpha\beta}a^\beta-\varepsilon_{\alpha\delta\kappa}
n^{\kappa\tau}\sigma^\delta{}_\tau = 0 \,. \label{cons}
\end{equation}
To find systems of equations for Bianchi models similar to those described 
in the FLRW case, we need evolution equations for the new variables:
$a_\alpha\,,$ $n^{\alpha\beta}$ and $\sigma_{\alpha\beta}\,.$  
However, it is better to consider them for each particular 
Bianchi case. 

\subsection{Bianchi I perfect-fluid cosmologies}
The Bianchi I models are homogeneous and anisotropic cosmological
models containing the flat FLRW spacetimes.  We can specialize
the triad $\{\mb{e}_\alpha\}$ in such a way that the unit vector
fields $\mb{e}_\alpha$ are Fermi-Walker propagated along $\mb{u}$ and at 
the same time their commutation functions vanish~\cite{EllMac:1969}
\begin{equation} 
\gamma^\delta{}_{\alpha\beta} = 0~\Leftrightarrow~a_\alpha =
n^{\alpha\beta} = 0 \,. \label{gaio}
\end{equation}
Then, in this case the constraint~(\ref{cons}) is identically satisfied.
Moreover, in these models the spatial curvature, the curvature of the
hypersurfaces orthogonal to the fluid velocity, vanishes, that is,
\[ {}^3R_{ab}=0 \,. \]
In particular ${}^3R=0$, which is a consequence of~(\ref{tcur}) 
and~(\ref{gaio}).  Then, in this case the Friedmann equation~(\ref{bfri}) 
takes the following form 
\begin{equation} 
\Omega_\rho + \Omega_\Lambda + \Omega_\sigma + \Omega_\lambda = 1 \,, 
\label{frbi}
\end{equation}
where $\Omega_\rho$, $\Omega_\Lambda$ and $\Omega_\lambda$ are defined
as in the FLRW case [see equations~(\ref{dlv1},\ref{dlv2})], and where
we have introduced the following dimensionless quantity associated with
the shear
\begin{equation} 
\Omega_\sigma\equiv \frac{\sigma^2}{3H^2} =
\frac{\sigma^{ab}\sigma_{ab}}{6H^2} \,. \label{dlv3}
\end{equation}
We can construct a state space for the Bianchi I cosmological models by
taking the variables $\mb{\Omega}=(\Omega_\rho,\Omega_\Lambda,\Omega_\sigma,
\Omega_\lambda)$.  Then, taking into account that all these quantities are 
positive by definition, the Friedmann equation~(\ref{frbi}) implies that we 
have got a compact state space in which these variable are restricted to 
the interval $[0,1]$.  Using the time derivative defined in~(\ref{tdkn}) 
and using the evolution equation for $\sigma^2$~\cite{EllMac:1969}
\begin{equation} 
\left(\sigma^2\right)^\cdot = - 6H\sigma^2 \,, \label{esic}
\end{equation}
the system of dynamical equations is given by
\begin{eqnarray}
\Omega'_\rho & = & \epsilon [2(1+q)-3\gamma]\Omega_\rho \,, \label{dbii} \\
\Omega'_\Lambda & = & 2\epsilon(1+q)\Omega_\Lambda \,, \\
\Omega'_\sigma & = & 2\epsilon(q-2)\Omega_\sigma\,, \\
\Omega'_\lambda & = & 2\epsilon[1+q-3\gamma]\Omega_\lambda\,, \label{dbif}
\end{eqnarray}
and the equation~(\ref{dece}), which again is uncoupled to the rest
of equations.  Now, the expression for the deceleration parameter $q$
in terms of the variables $\mb{\Omega}$ is given by
\begin{equation} 
q = \frac{3\gamma-2}{2}\Omega_\rho-\Omega_\Lambda+2\Omega_\sigma
+(3\gamma-1)\Omega_\lambda \,. \label{bdec}
\end{equation}
The critical points of the dynamical system~(\ref{dbii}-\ref{dbif}) having
a hyperbolic character, together with their state space coordinates
$\mb{\Omega}^\ast=(\Omega_\rho^\ast,\Omega_\Lambda^\ast,\Omega_\sigma^\ast,
\Omega_\lambda^\ast)$ and their eigenvalues are given in the following table:
\begin{quasitable}
\begin{tabular}{ccc}
Model & Coordinates & Eigenvalues  \\ \tableline
$\mbox{F}_\epsilon$ & $(1,0,0,0)$ & $\epsilon(3\gamma-2,3\gamma,
3(\gamma-2),-3\gamma)$ \\
$\mbox{dS}_\epsilon$ & $(0,1,0,0)$ & $-\epsilon(3\gamma,2,6,6\gamma)$  \\
$\mbox{K}_\epsilon$ & $(0,0,1,0)$ & $\epsilon(-3(\gamma-2),6,4,
-6(\gamma-1))$ \\
$\mbox{m}_\epsilon$ & $(0,0,0,1)$ & $\epsilon(3\gamma,6\gamma,6(\gamma-1),
2(3\gamma-1))$
\end{tabular}
\end{quasitable}
where $\mbox{K}$ denotes the Kasner vacuum spacetimes, whose line element
can be written as follows
\begin{equation} 
\mbox{ds}^2= -dt^2 + \sum_{\alpha=1}^3 t^{2p_\alpha} (dx^\alpha)^2 \,, 
\label{kasn}
\end{equation}
where $p_\alpha$ are constants satisfying
\begin{equation} 
\sum_{\alpha=1}^3 p_\alpha =\sum_{\alpha=1}^3 p^2_\alpha = 1\,. \label{kcon}
\end{equation}
Apart from the critical points shown in the table above, we have found
sets of infinite points in the particular cases $\gamma=0,1,2\,,$ which
at the same time constitute bifurcations and will be discussed later.
From the eigenvalues we get the dynamical
character of the critical points, which is shown in the table below
\begin{quasitable}
\begin{tabular}{cccc}
Model  & \multicolumn{3}{c}{Dynamical character} \\
\mbox{} &  $0<\gamma < 1$ & $\gamma=1$ & $\gamma > 1$  \\ \tableline
$\mbox{F}_{\pm}$ & saddle & saddle & saddle  \\
$\mbox{dS}_+$ & attractor & attractor & attractor  \\
$\mbox{dS}_-$ & repeller & repeller & repeller \\
$\mbox{K}_+$  & repeller & repeller & saddle  \\
$\mbox{K}_-$  & attractor & attractor & saddle  \\
$\mbox{m}_+$  & saddle & repeller & repeller  \\
$\mbox{m}_-$  & saddle & attractor & attractor
\end{tabular}
\end{quasitable}

Now let us analyze the state space for these models.  It can be
represented by the same drawings used for the ${}^3R\leq 0$ sector
of the FLRW evolution, the only thing we need to change is the
axis corresponding to the variable $\Omega_k$.  For the Bianchi I
models instead of $\Omega_k$ we have to consider $\Omega_\sigma$,
and instead of the critical points $\mbox{M}_\pm$ (Milne), we have to
consider $\mbox{K}_\pm$ (Kasner).  Then, taking into account this
correspondence, let us examine the structure of the state space for
Bianchi I models for the different values of $\gamma$. For the
sake of simplicity, we will do it only for expanding models, that is
$\epsilon=1$.  The case $\epsilon=-1$ can be obtaining by a simple
time reversal.

\begin{figure}
\epsfig{figure=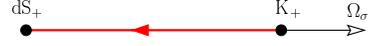,
height=0.5in,width=3in, bbllx=55, bblly=350, bburx=540, bbury=430}
\caption{State space for Bianchi I models with $\gamma=0\,.$} \label{big0}
\end{figure}

In the case $\gamma=0$, all the points given by $\mb{\Omega}^\ast
=(\Omega^\ast_\rho,\Omega^\ast_\Lambda,0,\Omega^\ast_\lambda)$ such
that $\Omega^\ast_\rho+\Omega^\ast_\Lambda+\Omega^\ast_\lambda=1$, are
critical points corresponding to the de Sitter model as given in
expressions~(\ref{dscl},\ref{mcco}) for $k=0$. The dynamics can be 
shown in an one-dimensional state space (see Figure~\ref{big0}).  
De Sitter is the attractor and Kasner the repeller, therefore, any 
initial anisotropy is diluted out in the evolution.

The situation in the case $\gamma\in(0,1)$ is more complicated.
The state space is represented in Figure~\ref{esp1}(b).
As we can see, $\mbox{F}_+$ is the attractor in the case without
cosmological constant.  When a cosmological constant is present, the 
de Sitter is the general attractor.  In both cases, the anisotropic 
Kasner models are repellers
of the evolution, which means that independently of the initial conditions
the models isotropize.

When $\gamma=1$, which corresponds to dust matter, a bifurcation occurs
and the state space is now given in Figure~\ref{esp2}(b).  The dynamical
behaviour is essentially the same as in the previous case, the difference
is that now we have an infinite set of critical points situated in the
line $\Omega_\sigma+\Omega_\lambda=1\,.$  We can find these models
by using the limiting procedure introduced in~(\ref{thel}). The line element 
of these models is described by the Kasner metric~(\ref{kasn}), but now 
only the first condition in~(\ref{kcon}) holds, i.e., 
\begin{equation} 
\sum_{\alpha=1}^3 p_\alpha = 1 \,, \label{only}
\end{equation}
where the parameters $p_\alpha$ depend on $\kappa_{(5)}$, and in the limit
$\rho\rightarrow 0$, where the influence of the extra dimension disappears, 
we recover the vacuum Kasner models.

In the interval $\gamma\in(1,2)$ the only change with respect to the
situation in the case $\gamma\in(0,1)$ is that now the BDL
model ($\mbox{m}_+$) is a repeller point and $\mbox{K}_{\pm}$ are
saddle points. The state space is represented in Figure~\ref{esp3}(b).
In $\gamma=2$, which corresponds to a stiff matter equation of state
($p=\rho$), there is another bifurcation.  We have a line of general
relativistic critical points, as shown by Figure~\ref{esp4}(b).
The models are described, as in the previous case ($\gamma=1$),
by a Kasner metric~(\ref{kasn}) where the parameters only 
satisfy~(\ref{only}) and the energy density is given by 
\begin{equation} 
\kappa^2\rho = \frac{1}{2t^2}\left(1-\sum_{\alpha=1}^3 p^2_\alpha\right) 
\label{edsm}\,. 
\end{equation}
They are saddle points.  The dynamics of the rest of the state space
is as in the case $\gamma\in(1,2)$.  Finally, for $\gamma> 2$
(which does not satisfy the causality condition), the state space
would be given by Figure~\ref{esp5}(b).

To sum up, we have seen that expanding models isotropize as it 
happens in general relativity, although now we can have 
intermediate stages in which the anisotropy can grow [see 
Figure~\ref{esp4}(b) for $\gamma\in(1,2)$]. The situation near
the Big Bang is more interesting.  In the brane-world scenario
anisotropy dominates only for $\gamma<1$, whereas in general
relativity it dominates for $\gamma<2$, therefore
in the physically relevant interval $\gamma\in(1,2)$ the prediction
is completely different (for the context of inflation 
see~\cite{MASS}): in the brane-world scenario the singularity is
isotropic.

\subsection{Bianchi V perfect-fluid cosmologies}
In the Bianchi V cosmological models the hypersurfaces of homogeneity,
which we have assumed to be orthogonal to the fluid velocity, are
negatively curved.  In fact, we can pick up a triad $\{\mb{e}_\alpha\}$ 
Fermi-Walker propagated along $\mb{u}$ and such that 
the spatial commutation functions satisfy 
\begin{equation} 
a_1\neq 0\,,~~~ a_2=a_3=0\,,~~~\mbox{and}~~~ 
n_{\alpha\beta}=0\,. \label{bvco}
\end{equation}
Then, the spatial scalar curvature~(\ref{tcur}) is given by
\[ {}^3R = -6a^2_1 < 0 \,. \]
We can introduce the quantity $\Omega_k$ as defined 
in~(\ref{dlv1}), but now it looks as follows
\[ \Omega_k\equiv -\frac{{}^3R}{6H^2}= \frac{a^2_1}{H^2} > 0 \,. \]
Therefore, using also the variables $\Omega_\rho$, $\Omega_\Lambda$,
$\Omega_\lambda$ [equations~(\ref{dlv1},\ref{dlv2})], and 
$\Omega_\sigma$~(\ref{dlv3}), the Friedmann equation~(\ref{bfri})
becomes
\[ \Omega_\rho+\Omega_k+\Omega_\Lambda+\Omega_\sigma+\Omega_\lambda=1\,.\]
This equation implies that we can construct a compact state space from the 
variables $\mb{\Omega}=(\Omega_\rho,\Omega_k,
\Omega_\Lambda,\Omega_\sigma,\Omega_\lambda)$, which are all positive,
and as usual, restricted to the interval $[0,1]$.

Before looking at the dynamical system for $\mb{\Omega}$ let us consider
the constraint~(\ref{cons}).  In this case it is not automatically
satisfied, but it imposes, by virtue of~(\ref{bvco}), the following 
condition
\begin{equation}  
\sigma^{}_{1\alpha} = 0 \,, \label{rest}
\end{equation}
which supposes a restriction on the general metric of the Bianchi V models,
whose line element can be written in the following way
\[ \mbox{ds}^2 = -dt^2+A^2(t)dx^2+\mbox{e}^{2x}\left[B^2(t)dy^2+
C^2(t)dz^2\right] \,. \]
The restriction imposed by~(\ref{rest}) is then
\begin{equation} 
A^2 = BC\,. \label{reco}
\end{equation}

To find the dynamical system we need the evolution equations for 
$a_1$ and $\sigma^2$.  The equation for $a_1$ is~\cite{EllMac:1969}
\[ \dot{a}{}_1 = -Ha_1\,, \]
and the equation for $\sigma^2$ is~(\ref{esic}).  Then, the equations
for $\mb{\Omega}$ are
\begin{eqnarray}
\Omega'_\rho & = & \epsilon [2(1+q)-3\gamma]\Omega_\rho \,, \label{dbvi} \\
\Omega'_k & = & 2\epsilon q\Omega_k \,, \\
\Omega'_\Lambda & = & 2\epsilon(1+q)\Omega_\Lambda \,, \\
\Omega'_\sigma & = & 2\epsilon(q-2)\Omega_\sigma\,, \\
\Omega'_\lambda & = & 2\epsilon[1+q-3\gamma]\Omega_\lambda\,, \label{dbvf}
\end{eqnarray}
where the deceleration parameter is also given by the expression~(\ref{bdec}).
The critical points of the dynamical system~(\ref{dbvi}-\ref{dbvf})
as well as their coordinates and eigenvalues are given in the following
table~\cite{Nota}
\begin{quasitable}
\begin{tabular}{ccc}
Model & Coordinates & Eigenvalues  \\ \tableline
$\mbox{F}_\epsilon$ & $(1,0,0,0,0)$ & $\epsilon(3\gamma-2,3\gamma-2,3\gamma,
3(\gamma-2),-3\gamma)$ \\
$\mbox{M}_\epsilon$ & $(0,1,0,0,0)$ & $-\epsilon(3\gamma-2,0,-2,4,
2(3\gamma-1))$ \\
$\mbox{dS}_\epsilon$ & $(0,0,1,0,0)$ & $-\epsilon(3\gamma,2,2,6,6\gamma)$  \\
$\mbox{K}^\ast_\epsilon$ & $(0,0,0,1,0)$ & $\epsilon(-3(\gamma-2),4,6,4,
-6(\gamma-1))$ \\
$\mbox{m}_\epsilon$ & $(0,0,0,0,1)$ & $2\epsilon(\textstyle{3\gamma\over2},
3\gamma-1,3\gamma,3(\gamma-1),3\gamma-1)$
\end{tabular}
\end{quasitable}
That is, we recover the equilibrium points we had in the case of FLRW
models with ${}^3R<0$ plus the models denoted by $\mbox{K}^\ast$, which
corresponds to Kasner models.  However, we must take into 
account the restriction~(\ref{reco}), which implies that the critical 
points $\mbox{K}^\ast$ only represent Kasner models for which the parameters 
$p_\alpha$ are given by
\begin{equation} 
p_1 = \frac{1}{3}\,,~~~ p_2 = \frac{1+\sqrt{3}}{3}\,,~~~ 
p_3 = \frac{1-\sqrt{3}}{3} \,. \label{expo}
\end{equation}
On the other hand, we have now sets of infinite points for 
$\gamma=0,\textstyle{1\over3},\textstyle{2\over3},1,2$, which also are  
bifurcation values of the parameter $\gamma$.  We have more 
bifurcations than in the Bianchi
I case, so we need more state space diagrams to represent the dynamics.
To do that we need to extract, from the previous table, the dynamical 
character of the equilibrium points, which is shown in the next table
\begin{quasitable}
\begin{tabular}{cccc}
Model  & \multicolumn{3}{c}{Dynamical character} \\
\mbox{} & $0<\gamma < 1$ & $\gamma=1$ & $\gamma > 1$  \\ \tableline
$\mbox{F}_\pm$ & saddle & saddle & saddle  \\
$\mbox{M}_\pm$ & saddle & saddle & saddle  \\
$\mbox{dS}_+$ & attractor & attractor & attractor  \\
$\mbox{dS}_-$ & repeller & repeller & repeller \\
$\mbox{K}^\ast_+$  & repeller & repeller & saddle  \\
$\mbox{K}^\ast_-$  & attractor & attractor & saddle  \\
$\mbox{m}_+$  & saddle & repeller  & repeller  \\
$\mbox{m}_-$  & saddle & attractor & attractor
\end{tabular}
\end{quasitable}

\begin{figure}
\begin{center}
\includegraphics[height=0.20in,width=2.4in]{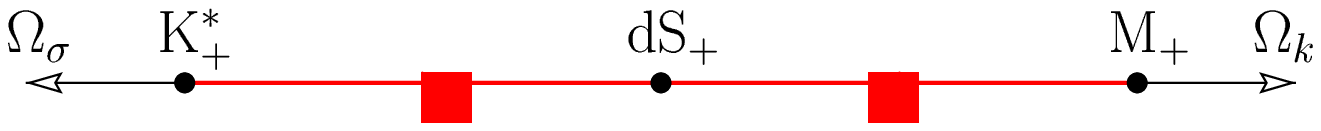}
\caption{State space for Bianchi V models with $\gamma=0\,.$}\label{fbv1}
\end{center}
\end{figure}

\begin{figure}
\begin{center}
\includegraphics[height=3in,width=3in]{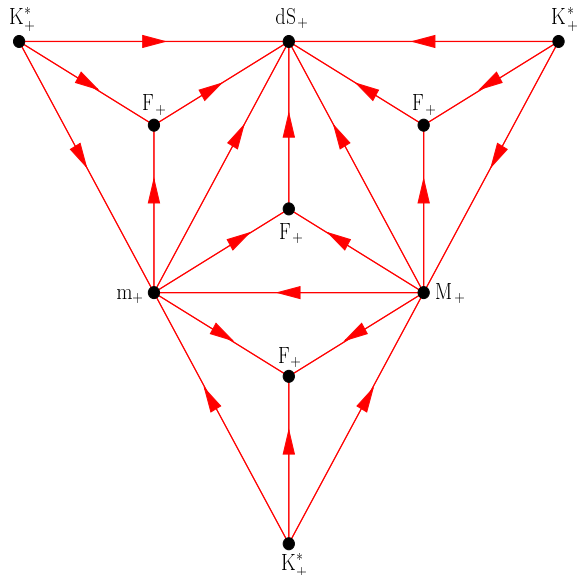}
\caption{State space for Bianchi V models with $\gamma\in(0,
\textstyle{1\over3})\,.$  Changing the line joining $\mbox{m}_+$ and
$\mbox{M}_+$ by a line in which all the points are critical we get the
state space for $\gamma=\textstyle{1\over3}\,.$ Reversing the 
arrow in that line we have the state space for $\gamma\in
(\textstyle{1\over3},\textstyle{2\over3})\,.$}\label{fbv2}
\end{center}
\end{figure}

\begin{figure}
\begin{center}
\includegraphics[height=3in,width=3in]{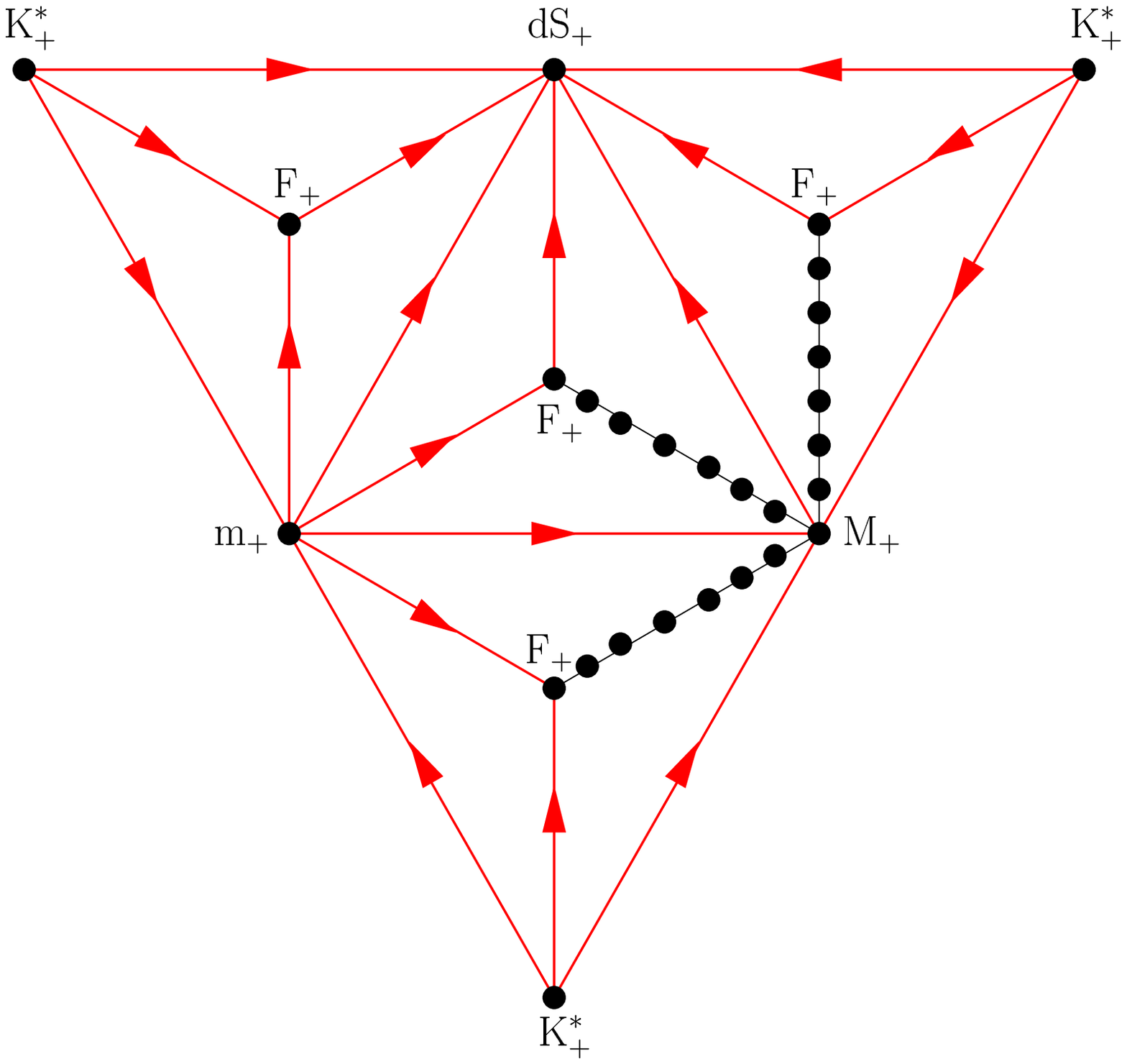}
\caption{State space for Bianchi V models with 
$\gamma=\textstyle{2\over3}\,.$}\label{fbv5}
\end{center}
\end{figure}

\begin{figure}
\begin{center}
\includegraphics[height=3in,width=3in]{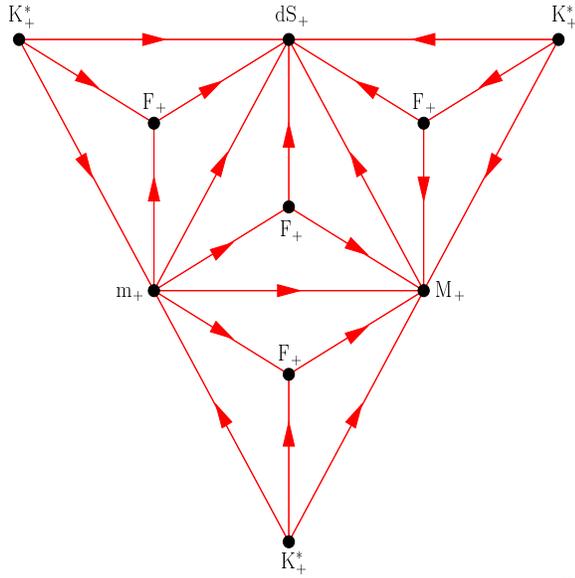}
\caption{State space for Bianchi V models with 
$\gamma\in(\textstyle{2\over3},1)\,.$ Changing the lines joining 
$\mbox{m}_+$ and $\mbox{K}^\ast_+$ by lines in which all the points are 
critical we get the state space for $\gamma=1\,.$ Reversing the 
arrow in those lines we have the state space for $\gamma\in(1,2)\,.$}
\label{fbv6}
\end{center}
\end{figure}

\begin{figure}
\begin{center}
\includegraphics[height=3in,width=3in]{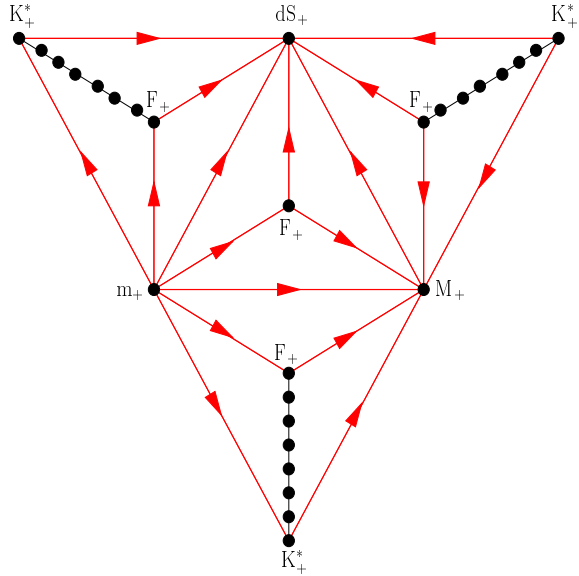}
\caption{State space for Bianchi V models with $\gamma=2\,.$}\label{fbv9}
\end{center}
\end{figure}

Let us now analyze the state space diagrams shown in Figures
\ref{fbv1}-\ref{fbv9}.  First of all we notice that to get diagrams 
similar to those of the FLRW and Bianchi I models we would need 
four-dimensional diagrams. However, this is not necessary since,
as it happens in the case of the FLRW and Bianchi I models, the 
qualitative dynamics follows from the trajectories of two-dimensional
invariant submanifolds.  Hence, we have drawn two-dimensional
state space diagrams in which all the dynamical information is 
present.  We have drawn only the trajectories joining critical
points and the direction of the dynamical flow. The interior trajectories
can be derived from them and by comparison with the state
space diagrams for the FLRW and Bianchi I models.             

We start with the case $\gamma=0$, in which the 
equation~(\ref{emce}) implies that the energy density is constant.  
The state space is again very simple (see Figure~\ref{fbv1}), the 
de Sitter model, with a modified cosmological constant~(\ref{mcco}), 
is the general attractor and the Kasner and Milne spacetimes are  
repellers.

Now, let us consider the case 
$\gamma\in(0,\textstyle{1\over3})\,,$ which is represented in  
Figure~\ref{fbv2}. Again, for the sake of brevity we will consider only the
expanding case, i.e., $\epsilon=1$ (the case $\epsilon=-1$
is obtained by time reversal, i.e. by reversing the arrow in the
state-space diagrams).  The attractor for $\Lambda=0$ is the
flat Friedmann model, whereas for $\Lambda\neq 0$ is de Sitter.  The
other critical points are saddle points.  As in the Bianchi I case, these
models isotropize (evolving either to $\mbox{F}_+$ or $\mbox{dS}_+$), with 
the exception of extreme situations of zero measure, representing models 
evolving towards the BDL solution.  
For $\gamma=\textstyle{1\over3}$ we have the first bifurcation due to
the appearance of a line with an infinite number of critical points located 
at $\mb{\Omega^\ast}=(0,\Omega_k^\ast,0,0,\Omega_\lambda^\ast)$ with
$\Omega_k^\ast+\Omega_\lambda^\ast=1$ (see Figure~\ref{fbv2}).  They
are the models discussed in the FLRW case (see Figure~\ref{esp2}).
In the case $\gamma\in (\textstyle{1\over3},\textstyle{2\over3})$ the only 
change with respect to the case $\gamma\in (0,\textstyle{1\over3})$ is
that now models in the line joining $\mbox{m}_+$ and $\mbox{M}_+$ evolve
to $\mbox{M}_+$ instead of $\mbox{m}_+\,.$ 

For $\gamma=2/3$ we have another bifurcation 
(see Figure~\ref{fbv5}) with a set of equilibrium points located
at $(\Omega^\ast_\rho,\Omega_k^\ast,0,0,0)$, with 
$\Omega_\rho^\ast+\Omega_k^\ast=1\,.$  They are FLRW models
that correspond to the critical points in Figure~\ref{esp4}(b).

In the interval $\gamma\in(\textstyle{2\over3},1)$ (Figure~\ref{fbv6})
we have that the flat FLRW model is a saddle point, even if we restrict
ourselves to the plane $\Omega_\Lambda=0\,.$ 
For $\gamma=1$ we have a bifurcation characterized by the appearance
of a set of anisotropic critical points which are not general relativistic
in nature since their coordinates are 
$(0,0,0,\Omega^\ast_\sigma,\Omega^\ast_\lambda)$ with
$\Omega^\ast_\sigma+\Omega^\ast_\lambda=1\,.$ They are among the critical 
points discussed in the case $\gamma=1$ of Bianchi I. The metric is
given by the line element~(\ref{kasn}), but now the exponents $p_\alpha$
satisfy  
\begin{equation}
p_1=\frac{1}{3}~~~\mbox{and}~~~p_2+p_3=\frac{2}{3}\,. \label{pote}
\end{equation}
In the situation $\gamma\in(1,2)$ (Figure~\ref{fbv6}), the only change 
with respect to the case $\gamma\in(\textstyle{2\over3},1)$ is that
the points $\mbox{K}^\ast_+$ are saddle instead of repellers.  This
means that if we initially start with 
$\Omega_\sigma^\ast+\Omega_\lambda^\ast=1\,,$ the models will evolve 
towards $\mbox{K}^\ast_+$ instead of $\mbox{m}_+\,.$

Finally, for $\gamma=2$ we have again an infinite number of critical points 
(see Figure~\ref{fbv9}). They are spatially-flat models and therefore, 
following the discussion of the Bianchi I models, they are described by a 
Kasner metric~(\ref{kasn}) with exponents given by~(\ref{pote}) and energy 
density by~(\ref{edsm}).

To summarize, we can say that the dynamics of the Bianchi V models 
encompasses the features of the state spaces of the FLRW models with 
${}^3 R\leq 0$ and the Bianchi I models.



\section{Remarks and conclusions}\label{core}
In this paper we have studied systematically the dynamics of  
homogeneous cosmological models (the FLRW and Bianchi I and V models)
in a generalized version of the scenario proposed by Randall and 
Sundrum~\cite{RanSun:1999b}, stressing the main differences with
respect to the general-relativistic case. 
In the case of the FLRW cosmological models, the state space presents
a new equilibrium point, namely, the BDL model 
($\mbox{m}_\pm$)~\cite{BinDefLan:2000}.   It dominates the dynamics 
at high energies, where the extra-dimension effects become dominant.
For this reason, we expect them to be a generic feature of the state
space of more general cosmological models in the brane-world scenario,
as it occurs in the Bianchi models analyzed here.  In the FLRW 
case the critical points $\mbox{m}_\pm$ describe the new dynamics near 
the Big Bang and also near the Big Crunch for recollapsing models.  

Another new feature is the existence of new bifurcations
as we change the equation of state, the parameter $\gamma\,.$
In the case of FLRW models there is one new bifurcation for 
$\gamma=\textstyle{1\over3}\,,$ characterized by the appearance
of an infinite number of non-general-relativistic critical points.
Among them we find a static model whose line-element is that of 
the Einstein universe.  This contrasts with general relativity,
where it appears for $\gamma=\textstyle{2\over3}\,.$  The consequence
is that in the brane-world scenario recollapsing models appear 
for $\gamma>\textstyle{1\over3}$ instead of $\gamma>\textstyle{2\over3}$
as in general relativity.  In the case of Bianchi I models we have
found a new bifurcation for $\gamma=1$, and in the case of Bianchi V
models for $\gamma=\textstyle{1\over3},1\,.$

On the other hand, Bianchi models allow us to study anisotropy.  We have
seen that expanding Bianchi I and V models always isotropize, as it happens
in general relativity, although now we can have intermediate stages in
which the anisotropy grows.  This is an expected result since the energy
density decreases and hence, the effect of the extra dimension becomes
less and less important. The situation changes drastically when we look
backwards.  Near the Big Bang the anisotropy only dominates for 
$\gamma\leq 1$, whereas in general relativity it dominates for
$\gamma<2$, which includes all the physically interesting cases.

Just to finish we would like to mention some current and future work 
in the line of the present one.  First we recall that in this paper we 
have considered brane-world scenarios in which the bulk 
satisfies the condition~(\ref{assu}).  Then, it would be interesting to 
look at the effect of having a contribution from the bulk curvature, or 
in other words, a contribution from the bulk Weyl tensor piece 
$E^{(5)}_{ab}$.  This is currently under investigation~\cite{CamSop:2000}.
On the other hand, taking into account that string theory is
formulated in spacetimes with more than one extra dimension (brane world
of codimension greater than one)
it would be interesting to study how the introduction of more
extra dimensions changes the results presented here.  In this sense,
a good starting point would be to consider scenarios like those
introduced in~\cite{ARDI}.



\[ \]
{\bf Acknowledgments:}
We would like to thank Roy Maartens for helpful comments and discussions.  
This work has been supported by the European Commission (contracts 
HPMF-CT-1999-00149 and HPMF-CT-1999-00158).




\begin{thebibliography}{10}

\bibitem{RanSun:1999b}
L.~Randall and R.~Sundrum, Phys. Rev. Lett {\bf 83},  4690  (1999).

\bibitem{LOWD}
V.~A.~Rubakov and M.~E.~Shaposhnikov, Phys. Lett. B {\bf 125},  136  (1983);
M.~Visser, {\em ibid.} {\bf 159},  22  (1985);
E.~J. Squires, {\em ibid.} {\bf 167},  286  (1986);
M.~Gell-Mann and B.~Zwiebach, Nucl. Phys. {\bf B260},  569  (1985);
H.~Nicolai and C.~Wetterich, Phys. Lett. B {\bf 150},  347  (1985);
M.~Gogberashvili, Mod. Phys. Lett. A {\bf 14},  2025  (1999).
K.~Akama, in {\em Lectures in Physics}, Vol. 176, edited by K.~Kikkawa,
N.~Nakanishi, and H.~Nariai (Springer Verlag, New York, 1982).

\bibitem{HorWit:1996a}
P.~Ho$\check{\mbox{r}}$ava and E.~Witten, Nucl. Phys. {\bf B460},  506  (1996).

\bibitem{HorWit:1996b}
P.~Ho$\check{\mbox{r}}$ava and E.~Witten, Nucl. Phys. {\bf B475},  94  (1996).

\bibitem{HOWI}
K.~Benakli, Int. J. Mod. Phys. D {\bf 8},  153  (1999);
A.~Lukas, B.~A.~Ovrut, K.~S.~Stelle, and D.~Waldram, Phys. Rev. D {\bf 59},
  086001  (1999);
H.~S.~Reall, {\em ibid.} {\bf 59},  103506  (1999);
A.~Lukas, B.~A.~Ovrut, and D.~Waldram, {\em ibid.} {\bf 60},  086001  (1999);
A.~Lukas, B.~A.~Ovrut, and D.~Waldram, {\em ibid.} {\bf 61},  023506  (1999);
A.~P.~Billyard, A.~A.~Coley, J.~E.~Lidsey, and U.~S.~Nilsson,  {\em ibid.}
{\bf  61},  043504  (2000).

\bibitem{RanSun:1999a}
L.~Randall and R.~Sundrum, Phys. Rev. Lett {\bf 83},  3370  (1999).

\bibitem{UNFO}
C.~Cs$\acute{\mbox{a}}$ki, M.~Graesser, C.~Kolda, and J.~Terning, 
Phys. Lett. B  {\bf 462},  34  (1999);
J.~M.~Cline, C.~Grojean, and G.~Servant, Phys. Rev. Lett. {\bf 83},  4245
  (1999).

\bibitem{BinDefLan:2000}
P.~Bin$\acute{\mbox{e}}$truy, C.~Deffayet, and D.~Langlois, Nucl. Phys. {\bf
  B565},  269  (2000).

\bibitem{BinDefEllLan:2000}
P.~Bin$\acute{\mbox{e}}$truy, C.~Deffayet, U.~Ellwanger, and D.~Langlois, Phys.
  Lett. B {\bf 477},  285  (2000).

\bibitem{FlaTyeWas:2000b}
$\acute{\mbox{E}}$.~$\acute{\mbox{E}}$.~Flanagan, S.-H.~H.~Tye, and 
I.~Wasserman, Phys. Rev. D {\bf 62},  044039  (2000).

\bibitem{EXDI}
N.~Arkani-Hamed, S.~Dimopoulos, and G.~Dvali, Phys. Lett. B {\bf 429},  263
  (1998);
I.~Antoniadis, N.~Arkani-Hamed, S.~Dimopoulos, and G.~Dvali, {\em ibid.} 
{\bf  436},  257  (1998);
N.~Arkani-Hamed, S.~Dimopoulos, and G.~Dvali, Phys. Rev. D {\bf 59},  086004
  (1999);
N.~Arkani-Hamed, L.~Hall, D.~Smith, and N.~Weiner, {\em ibid.} {\bf 62},
  105002  (2000);
N.~Kaloper, J.~March-Russell, G.~D.~Starkman, and M.~Trodden, Phys. Rev. Lett.
  {\bf 85},  928  (2000).

\bibitem{ChuEveDav:2000}
D.~J.~H.~Chung, H.~Davoudiasl, and L.~Everett, hep-ph/0010103  (2000).

\bibitem{MaaWanBasHea:2000}
R.~Maartens, D.~Wands, B.~A.~Bassett, and I.~P.~C.~Heard, Phys. Rev. D {\bf
  62},  041301  (2000).

\bibitem{LonChaPri:1999}
J.~C.~Long, H.~W.~Chan, and J.~C.~Price, Nucl. Phys. {\bf B539},  23  (1999).

\bibitem{HoyEotWas:2000}
C.~D.~Hoyle {\it et al}, Phys. Rev. Lett. {\bf 86}, 1418 (2001).

\bibitem{ShiMaeSas:2000}
T.~Shiromizu, K.~Maeda, and M.~Sasaki, Phys. Rev. D {\bf 62},  024012  (2000).

\bibitem{SasShiMae:2000}
M.~Sasaki, T.~Shiromizu, and K.~Maeda, Phys. Rev. D {\bf 62},  024008  (2000).

\bibitem{Maa:2000}
R.~Maartens, Phys. Rev. D {\bf 62},  084023  (2000).

\bibitem{Wal:1984}
R.~M.~Wald, {\em General Relativity} (The University of Chicago Press, 
Chicago, 1984).

\bibitem{CamSop:2000}
A.~Campos and C.~F.~Sopuerta, (in preparation).

\bibitem{DSCO}
J.~Wainwright and G.~F.~R.~Ellis, {\em Dynamical systems in cosmology}
  (Cambridge University Press, Cambridge, 1997);
A.~A.~Coley, in {\em Proceedings of the Spanish Relativity Meeting. ERE-99},
edited by J.~Ib{\'a}{\~n}ez (Euskal Herriko Unibertsitatea,
  Bilbo, 2000).

\bibitem{DSTH}
D.~K.~Arrowsmith and C.~M.~Place, {\em An introduction to dynamical systems}
  (Cambridge University Press, Cambridge, 1990);
D.~W.~Jordan and P.~Smith, {\em Nonlinear ordinary differential equations. An
  introduction to dynamical systems} (Oxford University Press, Oxford, 1999).

\bibitem{HawEll:1973}
S.~W.~Hawking and G.~F.~R.~Ellis, {\em The large scale structure of space-time}
  (Cambridge University Press, Cambridge, 1973).

\bibitem{REGE}
S.~Weinberg, {\em Gravitation and cosmology} (John Wiley \& Sons, New York,
  1972);
C.~W.~Misner, K.~S.~Thorne, and J.~A.~Wheeler, {\em Gravitation} 
(W.~H.~Freeman and Company, New York, 1973).

\bibitem{Rie:1998}
A.~G.~Riess {\it et al}, Astron. Journ. {\bf 116},  1009  (1998).

\bibitem{Per:1999}
S.~Perlmutter {\it et al}, Astrophys. Journ. {\bf 517},  565  (1999).

\bibitem{GolEll:1999}
M.~Goliath and G.~F.~R.~Ellis, Phys. Rev. D {\bf 60},  023502  (1999).

\bibitem{Nota} The presence of a zero eigenvalue is (excepting in the
case of the Einstein Universe critical points) a consequence of having
a dynamical system with a constraint: The Friedmann equation. If we had
used it to eliminate one variable and reduce the dimension of the state 
space, we would not find any vanishing eigenvalue.
In conclusion, this fact does not affect to the hyperbolic character of 
the critical points. 


\bibitem{EllMac:1969}
G.~F.~R.~Ellis and M.~A.~H.~MacCallum, Commun. Math. Phys. {\bf 12},  108
  (1969).

\bibitem{MASS} 
R.~Maartens, V.~Sahni and T.~D.~Saini, Phys. Rev. D {\bf 63}, 063509
(2001).

\bibitem{ARDI}
N.~Arkani-Hamed, S.~Dimopoulos, G.~Dvali, and N.~Kaloper, Phys. Rev. Lett. 
{\bf 84},  586  (2000);
C.~Csaki and Y.~Shirman, Phys. Rev. D {\bf 61}, 024008 (2000);
A.~E.~Nelson, {\em ibid.} {\bf 63}, 087503 (2001);
T.~Gherghetta and M.~Shaposhnikov, Phys. Rev. Lett. {\bf 85}, 240 (2000);
T.~Gherghetta, E.~Roesl and M.~Shaposhnikov, Phys. Lett. B {\bf 491},
353 (2000).

\end{thebibliography}


\end{document}